\documentclass[journal]{IEEEtran}
\usepackage{cite}
\usepackage{graphicx}
\usepackage{amsmath} 
\usepackage{amssymb} 
\usepackage{subfigure}  
\usepackage{multirow} 
\usepackage{enumerate}
\usepackage{amsthm} 

\begin{document}
\title{Optimum Transmission Window for EPONs with Gated-Limited Service}

\author{Huanhuan~Huang,
        Tong~Ye,~\IEEEmembership{Member,~IEEE,}
        Tony~T.~Lee,~\IEEEmembership{Fellow,~IEEE,}
        and~Weisheng~Hu,~\IEEEmembership{Member,~IEEE}
\thanks{This work was supported in part by the National Science Foundation of China under Grant 61571288, Grant 61671286, and Grant 61433009, and in part by the Open Research Fund of Key Laboratory of Optical Fiber Communications (Ministry of Education of China).} 
\thanks{The authors are with the State Key Laboratory of Advanced Optical Communication Systems and Networks, Shanghai Jiao Tong University, Shanghai 200240, China (email: huanghuanhuan@sjtu.edu.cn; yetong@sjtu.edu.cn; ttlee@sjtu.edu.cn; wshu@sjtu.edu.cn).}}

\markboth{Submitted to IEEE/ACM Transactions on Networking}%
{Huang \MakeLowercase{\textit{et al.}}: Optimum Transmission Window for EPONs with Gated-limited Service}
\maketitle

\begin{abstract}
This paper studies the Ethernet Passive Optical Network (EPON) with gated-limited service. The transmission window (TW) is limited in this system to guaranteeing a bounded delay experienced by disciplined users, and to constrain malicious users from monopolizing the transmission channel. Thus, selecting an appropriate TW size is critical to the performance of EPON with gated-limited service discipline. To investigate the impact of TW size on packet delay, we derive a generalized mean waiting time formula for M/G/1 queue with vacation times and gated-limited service discipline. A distinguished feature of this model is that there are two queues in the buffer of each optical network unit (ONU): one queue is inside the gate and the other one is outside the gate. Furthermore, based on the Chernoff bound of queue length, we provide a simple rule to determine an optimum TW size for gated-limited service EPONs. Analytic results reported in this paper are all verified by simulations.
\end{abstract}


\section{Introduction}
\IEEEPARstart{T}{he} ever-growing Internet traffic generated by emerging services, including video on demand, remote e-learning, and online gaming, continuously exacerbates the last mile bottleneck problem in recent decades \cite{Kramer2001}. Ethernet Passive Optical Network (EPON) has been considered as an attractive solution to this problem due to its low cost, large capacity and ease of upgrade to higher bit rates \cite{EPONbook}. It has been deployed widely in many access networks such as Fiber-To-The-Home (FTTH), Fiber-To-The-Building (FTTB) and Fiber-To-The-Curb (FTTC) \cite{ftth2005,fttx2006,fttx2008}.

\begin{figure}[!t]
\centering
\includegraphics[width=0.45\textwidth]{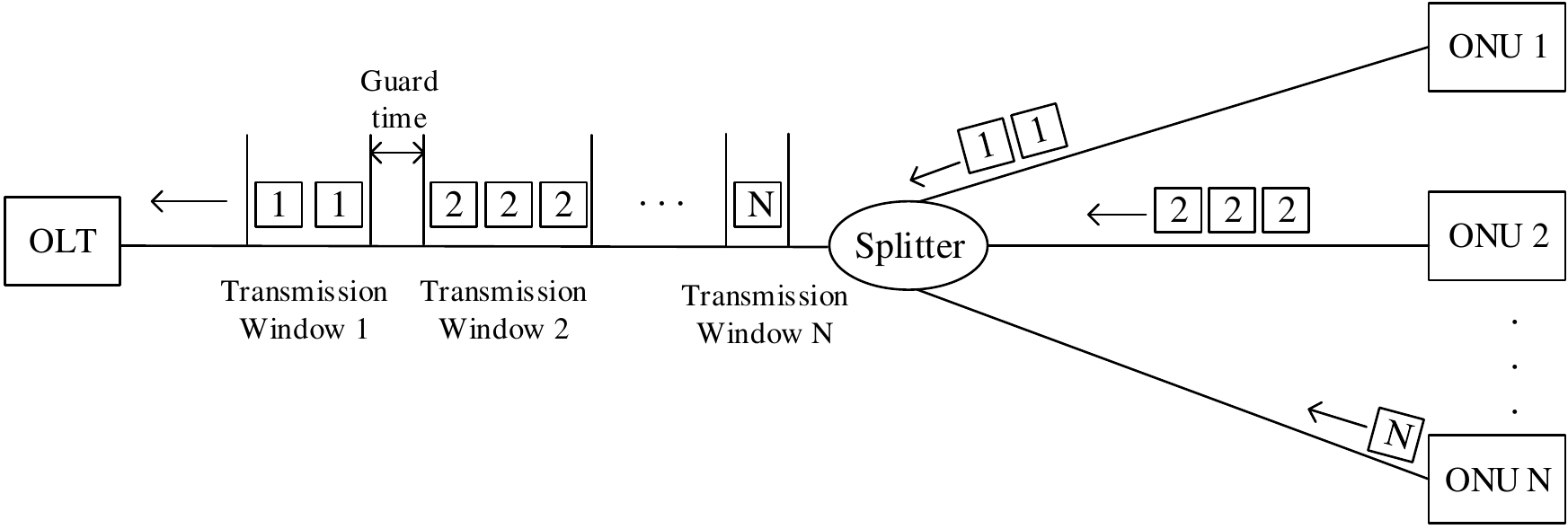}
\caption{Upstream transmission in the EPON.}
\label{fig1}
\end{figure}

A typical EPON is plotted in Fig.~\ref{fig1}. An EPON is a point-to-multipoint network, where one optical line terminal (OLT) in the central office is connected to multiple optical network units (ONUs) located at the users' premises via an optical passive splitter. In the downstream direction, the OLT broadcasts the packets to all the ONUs, and each ONU only accepts the packets destined to it. In the upstream direction, the OLT schedules the ONUs to share the bandwidth in a time division multiplexing (TDM) manner. The OLT assigns transmission windows (TWs) to each ONU through sending GATE messages in a round-robin fashion. Upon receiving the GATE message, the ONU transmits upstream data in the allocated TW. The number of packets that the ONU can send during a TW is called the TW size in this paper. After data transmission, the ONU generates a REPORT message to inform the OLT of its buffer status \cite{EPONbook}. The TWs of two successive ONUs are separated by a guard time to avoid data overlapping. The sizes of TWs that the OLT allocates to each ONU depend on the service discipline that the OLT adopts.

The gated service discipline has been widely studied in previous works \cite{BharatiThailand,gatedIPACT,AnalyticalModel,aurzada2008delay,quasi-leavedpolling,long-reachPON,meanvalue,symmetricpolling}. In the gated service, each ONU is authorized to transmit the amount of data that it requests in the REPORT \cite{kramer2002diffCoS}. Thus, the gated service may lead to the phenomenon called the ¡°capture effect¡± \cite{Bluetooth}, when an ONU with heavy traffic monopolizes the upstream channel for a long time and transmits excessive amounts of data. The capture effect will impose a large delay to other ONUs and thus impair the quality of service (QoS) of other ONUs. With the gated-limited service, the EPON users have to sign a service level agreement (SLA) with the network operator to specify the upstream traffic rate, and the OLT typically sets a limit of the maximum TW size to guarantee the QoS of each ONU according to their signed SLAs.

The selection of the maximum TW size is a critical choice in the gated-limited service EPON. On one hand, if the maximum TW size is set too small, the backlog of the ONUs cannot be cleaned up quickly and the upstream bandwidth is wasted by a large number of guard times and REPORT messages. In this case, the ONU will suffer from a large delay. On the other hand, if the maximum TW size is set too large, the capture effect cannot be suppressed effectively. An extreme case is that the gated-limited service discipline will change to the gated service discipline when the limit goes to infinity.

In existing literature, only a few previous works have studied the selection of the maximum TW size via simulations. The impact of the maximum TW size on delay performance of an ONU is discussed in \cite{kramer2002ipact}, in which the author points out that the maximum TW size for each ONU can be fixed based on the SLA, but doesn't provide any concrete scheme for the selection of the maximum TW size. The aim of our paper is to develop a systematic method to select a proper maximum TW size for gated-limited service EPONs.

The upstream transmission process of each ONU can be described by a vacation queuing system, in which each TW of the ONU is considered as a busy period while the time between two successive TWs of the ONU is treated as a vacation period. In general, the modeling of a vacation queuing system with limited service discipline is quite difficult \cite{state-dependentlimit}. In the gated-limited service EPON, the number of packets that an ONU can transmit in a TW is limited by the maximum TW size. Thus, before the transmission of an arrival packet, it may have to wait multiple vacations, which is typically difficult to analyze \cite{datanetwork}.

\subsection{Previous Works Related to EPONs with Gated-Limited Service}
The exhaustive type $k$-limited vacation queuing systems were studied in \cite{leeM/G/1/N,iterativek-limited,time-limitandk-limit,twoqueuelimit}, where the server takes a vacation when either a queue has been emptied or a predefined number of $k$ customers have been served during the visit. In \cite{leeM/G/1/N}, the distributions of queue length, waiting time and busy period were obtained by using the embedded Markov chain and a combination of the supplementary variables and sample biasing techniques. In \cite{iterativek-limited,time-limitandk-limit}, the authors used matrix-analytic techniques to iteratively calculate the queue length distribution. In \cite{twoqueuelimit}, a polling system with two priority queues and $k$-limited service discipline was analyzed, where the high priority queue is served with queue length dependent service time while the low priority queue is served with constant service time. The high priority queue length distribution at departure instants were derived by the embedded Markov chain. However, these models cannot be directly applied to gated-limited service EPONs, in which the OLT only serves the packets that arrived before the last REPORT message of an ONU up to a predefined number, regardless if the buffer is empty or not.

The gated type $k$-limited service vacation queuing systems were considered in \cite{TakagiQueueingBook,vacationmodel}, where the server serves at most $k$ customers that present at a queue upon visiting and then begins a vacation. A queuing model based on an embedded Markov chain was developed in \cite{TakagiQueueingBook,vacationmodel} to derive the Laplace-Stieltjes transforms of waiting time and busy period distributions, but the computation is too complex to give a clear physical insight into the performance of the entire system. To resolve this problem, a simple geometric approach was proposed in \cite{datanetwork} to obtain the mean waiting time, but this approach can only solve a special case when the user is allowed to transmit one packet in each busy period.

In exiting literature, a few works were devoted to the modeling of gated-limited service EPONs. In \cite{BharatiThailand}, the authors gave an approximate expression of mean waiting time for a gated-limited service (which is called limited service in \cite{BharatiThailand}) EPON under the assumption that the maximum TW size in terms of time (instead of the number of packets) is quite large, which is actually similar to analysis of the gated service EPON. In \cite{AnalyticalModel}, an approximate mean delay of gated-limited service EPONs is derived by using a discrete Markov chain, which is invalid when traffic load is high.

In summary, none of the previous works have obtained a useful formula of mean waiting time for general gated-limited service EPONs where the maximum TW size is finite and larger than one, and neither have they discussed how to select a proper maximum TW size for each ONU of gated-limited service EPONs.

\subsection{Our Approach and Contributions}
In this paper, we analyze the polling process of EPONs with gated-limited service discipline. Our goal is to develop an insightful model to describe the delay performance of gated-limited service EPONs, and to find a systematic method of selecting the maximum TW size for each ONU based on the SLA.

First, we adopt the geometric approach described in \cite{datanetwork} to derive the mean waiting time of an M/G/1 queue with vacations and gated-limited service. A key step is to calculate the mean number of whole vacations, excluding the residual vacation, experienced by an arrival before it receives service. The computation of this key parameter is based on an innovative approach that establishes the connection between the mean number of whole vacations and the first and second moments of the number of packets served in a busy period. A distinguished feature of this model is that there are two queues in the buffer of each ONU: one queue is inside the gate and the other one is outside the gate.

Next, we apply the Chernoff bound of queue length to select the optimum TW size. According to the SLA, the delay performance of an ONU shouldn't be influenced by the TW size limit if its traffic rate does not exceed the subscribed rate. Thus, the criterion of selecting the optimum TW size is to choose the smallest integer that makes the probability of queue length exceeding the TW size limit negligible. That is, when an ONU operates in the subscripted region, its buffer can be emptied with a high probability at the end of every busy period. Otherwise, the ONU will suffer from a large delay when the input traffic rate exceeds the subscribed rate. Our specific contributions are summarized as follows:
\begin{enumerate}[{1.}]
\item
We derive a generalized formula of mean waiting time for M/G/1 queue with vacation time and gated-limited service discipline, which includes the mean waiting times of two queues.
\item
We provide a simple rule to determine a proper optimum TW size for ONUs of the gated-limited service EPON based on their SLAs, which is proved to be effective by simulations.
\end{enumerate}

The remainder of this paper is organized as follows. In Section II, we demonstrate the ¡°capture effect¡± and provide an overview of the polling process between the OLT and ONUs of a gated-limited service EPON. In Section III, we derive the mean waiting time of the M/G/1 queue with vacation time and gated-limited service discipline, and apply the result to ONUs of the EPON. Section IV discusses the method of selecting the optimum TW size and the delay performance of gated-limited service EPONs under the selected TW size. Section V draws the conclusion.

\section{Motivation and Overview}
\begin{figure*}[!t]
\centering
\subfigure[]{
\label{fig2-a}
\includegraphics[width=0.4\textwidth]{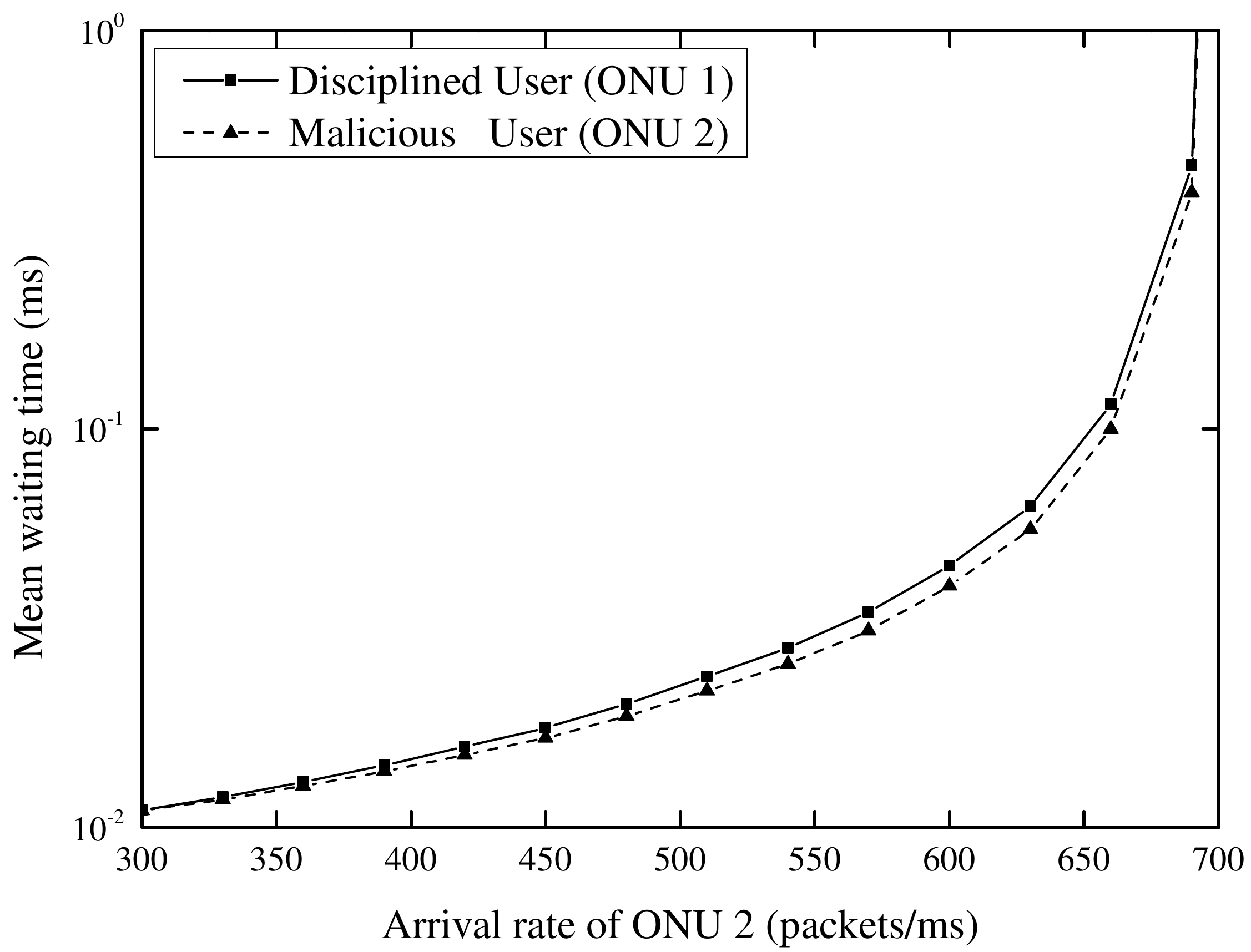}}
\subfigure[]{
\label{fig2-b}
\includegraphics[width=0.4\textwidth]{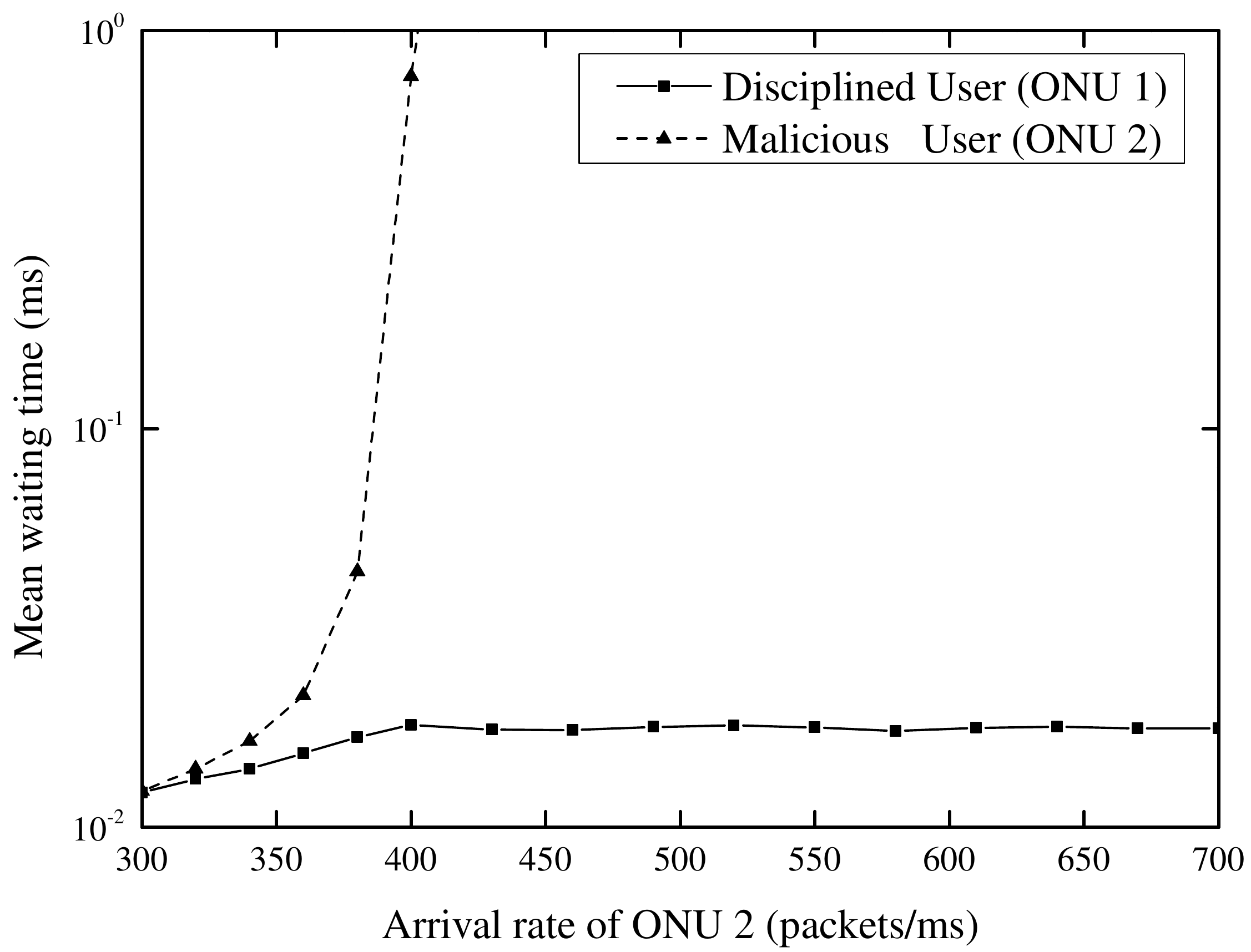}}
\caption{ Mean waiting time of two ONUs under different service disciplines: (a) gated service and (b) gated-limited service ($M=4$).}
\end{figure*}
An EPON can be considered as a polling system where a single server visits a set of queues in a cyclic order. A service discipline is one of the three policies, exhaustive service, gated service, and limited service, that specify the criteria of the server when progressing to the next queue \cite{takagipollingarticle}.
In a polling system with exhaustive service discipline where the server serves a queue until it becomes empty, the capture effect occurs when a heavily loaded user transmits excessive amounts of data and monopolizes the channel for a long time, such that other lightly loaded users suffer from prolonged waiting times.

To alleviate this problem, current EPON systems adopt the gated service discipline where the OLT only transmits the packets that are requested by the ONU in the last REPORT message during a TW. However, the capture effect still persists since the user with heavy traffic can report a large number of packets to the OLT at the end of a TW to secure a large size TW in the next cycle, which will lengthen the delays that other ONUs have to endure. This problem can be completely solved by the gated-limited service discipline, which is able to cap the TW size assigned to each ONU by a predefined value.

This point can be illustrated by using the example where two ONUs are connected to the OLT, and each of them is equipped with an infinite buffer. The service capacity of EPON is 1000 packets/ms. We assume that these two ONUs have signed the same SLAs with the network operator, meaning that they subscribe to the same upstream traffic rate (300 packets/ms) and have identical TW sizes, 4. Suppose ONU 1 is a disciplined user with a fixed input rate of 300 packets/ms according to the signed SLA, whereas ONU 2 is a malicious user whose traffic input rate is more than 300 packets/ms. Their mean waiting times versus the arrival rate of ONU 2 under the gated and gated-limited service disciplines are plotted in Fig.~\ref{fig2-a} and~\ref{fig2-b}, respectively.

With the gated service discipline, despite that the constant loading of ONU 1 is smaller than that of ONU 2, as Fig.~\ref{fig2-a} shows, the mean waiting time of ONU 1 is not only continuously increasing with the arrival rate of ONU 2, but also is uniformly larger than that of the malicious user ONU 2. Moreover, it is unbounded when the input traffic rate generated by ONU 2 reaches 700 packets/ms. In this case, 70 percent of bandwidth is monopolized by ONU 2.

On the other hand, with the gated-limited service discipline, each ONU can transmit no more than 4 packets during each busy period in our example. Fig.~\ref{fig2-b} shows that once the arrival rate of ONU 2 exceeds the subscribed rate 300 packets/ms, it immediately suffers from a larger mean waiting time than ONU 1 and soon becomes unstable when the arrival rate is larger than 400 packets/ms due to the limited TW size. The disciplined user ONU 1 enjoys a small mean waiting time all the time, and it is immune to the malicious behavior of ONU 2. This example clearly shows that the EPON with gated-limited service discipline can completely avoid the capture effect and provide a fair service to disciplined users while penalizing malicious users.

\begin{figure*}[!t]
\centering
\includegraphics[width=0.8\textwidth]{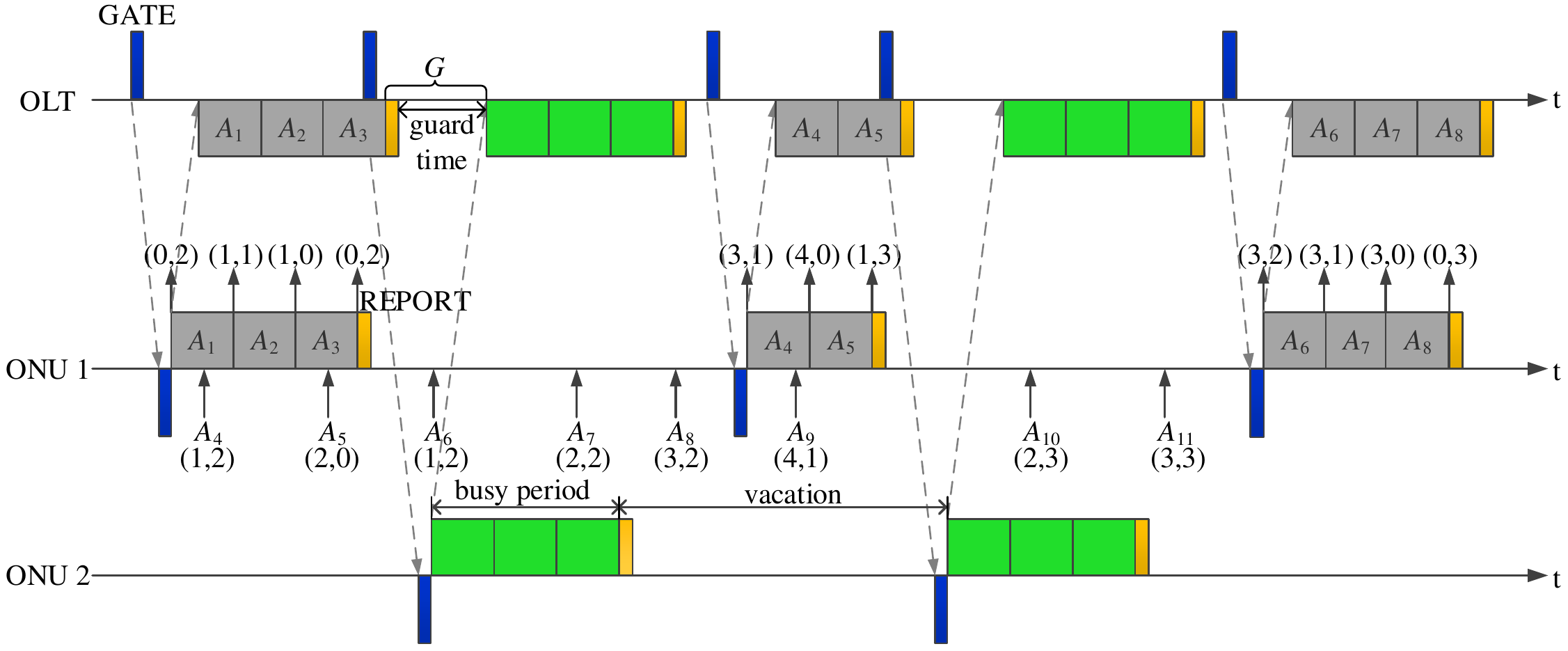}
\caption{Polling process of an EPON where $N=2$ and $M=3$.}
\label{fig3}
\end{figure*}

In an EPON system with $N$ ONUs that adopts the gated-limited service discipline, the packets waiting in the buffer of each ONU are divided into two groups by a fictitious gate. The number of packets inside the gate is bounded by the maximum TW size, denoted by $M$. An arrival packet first waits outside the gate and then enters the gate before it can be transmitted. As Fig.~\ref{fig3} illustrates, the buffer status is represented by a two-tuple state $(n,m)$, where $n$ is the number of packets waiting outside the gate, and $m$ is that waiting inside the gate. The number $n$ increases by 1 upon a new arrival, and $m$ decreases by 1 when a packet inside the gate begins to be transmitted by the ONU.

Fig.~\ref{fig3} plots the polling process of an EPON, where $N=2$ and $M=3$. A 64-byte GATE is employed by the OLT to notify an ONU about the start time and the length of each allocated TW. Upon receiving the GATE message, the ONU transmits all packets inside the gate during the TW. At the end of the TW, the ONU sends a 64-byte REPORT to the OLT, which reports the number of packets waiting outside the gate. According to the number $n$ stated in the REPORT, the OLT decides the TW size for this ONU in the next cycle, which equals the smaller of $M$ and $n$. Thus, the message REPORT offers the admission for packets waiting outside the gate to enter the gate.

After the ONU 1 issued the first REPORT, as Fig.~\ref{fig3} shows, the buffer state changes from $(2,0)$ to $(0,2)$, which means two packets entered the gate. Then the ONU 1 becomes idle while the OLT polls the next ONU. When the OLT finishes the polling of all other $(N-1)$ ONUs, it sends a GATE message to ONU 1 again to repeat the process. To avoid data overlapping induced by the clock synchronization problem between the OLT and the ONUs \cite{TONfittingreport}, two successive TWs are separated by a guard time. As Fig.~\ref{fig3} depicts, the constant interval $G$ includes the guard time and the transmission time of a REPORT.

The purpose of limiting the TW size is twofold: to guarantee a bounded delay experienced by disciplined users, and to constrain malicious users from monopolizing the transmission channel. With the gated-limited service discipline, the TW size $M$ limits the maximum number of packets that can be served in a busy period. The gated-limited service discipline cannot effectively constrain malicious users if the TW size $M$ is too large, and it may introduce longer than expected delays for disciplined users if $M$ is too small. Therefore, selecting an appropriate maximum TW size $M$ is critical to the performance of EPON with gated-limited service discipline. In the following sections, we devise a queueing model to analyze the mean waiting time of each ONU under the gated-limited service discipline, and provide a rule to determine the proper TW size $M$ of an EPON. The notations used throughout this paper are defined as follows for easy reference.
\begin{description}
\item[$N$]
Number of ONUs
\item[$M$]
The maximum TW size
\item[$G$]
The guard time plus the transmission time of a REPORT
\item[${\lambda}_{E}^{*}$]
Subscribed traffic rate for all the ONUs
\item[${\lambda}_{E}$]
Arrival rate of all the ONUs
\item[${\rho}_{E}$]
Offered load to the EPON
\item[$n_i$]
Number of packets found waiting outside the gate of the buffer when the $i$-th packet arrives
\item[$m_i$]
Number of packets found waiting inside the gate of the buffer when the $i$-th packet arrives
\item[$N_i$]
Number of packets found waiting in the buffer when the $i$-th packet arrives, ${N}_{i}={m}_{i}+{n}_{i}$
\item[$R_i$]
Residual time seen by the $i$-th packet when it arrives during a busy period or a vacation time in progress
\item[$Y_i$]
Duration of all the whole vacation times experienced by the $i$-th packet before it gets service
\item[$X_i$]
Service time of the $i$-th packet
\item[$W_i$]
Waiting time of the $i$-th packet
\end{description}

\section{Modeling of EPONs with Gated-limited Service}
In this section, we analyze the mean packet waiting time of an ONU in the EPON with gated-limited service discipline. As Fig.~\ref{fig3} shows, an ONU is busy with packet transmission during the TW, followed by a vacation period with a duration that equals the sum of $NG$ and the TWs of other $(N-1)$ ONUs. Therefore, at the end of each busy period, a predominate number of packets that an ONU reports to the OLT attributes to the number of arrivals during the vacation period before it. Hence, there are dependencies among TW sizes of ONUs. However, in the analysis of multiple access systems, such as Aloha or CSMA, this kind of dependency is weak and can be neglected when $N$ is large \cite{aloha,csma}. Thus, we can treat each ONU independently. We make the following assumptions in the modeling of EPONs with gated-limited service:
\begin{itemize}
\item[{A1.}]
All ONUs in the EPON are statistically identical.
\item[{A2.}]
The number of ONUs $N$ is large, such that the TWs of the ONUs can be considered as i.i.d. random variables. In practice, the number of ONUs is usually more than 16 in a typical EPON.
\item[{A3.}]
The packet arrival process of the EPON is Poisson, as is the arrival process of each ONU.
\item[{A4.}]
The packets are transmitted in a first-in-first-out (FIFO) manner, and the transmission times of the packets are i.i.d. random variables with a general distribution.
\item[{A5.}]
The propagation delay between the OLT and an ONU is very small in currently commercial EPONs, and can be ignored in the analysis.
\end{itemize}

Under these assumptions, each ONU can be considered as an M/G/1 queue with vacations and gated-limited service. In Section III-A, we derive the mean waiting time of this queuing system, which provides the key to analyze EPONs with gated-limited service discipline in this paper.

\subsection{M/G/1 Queue with Vacations and Gated-Limited Service}
We adopt the following notations in the analysis of the M/G/1 queue with vacations and gated-limited service:
\begin{enumerate}[{(1)}]
\item
the traffic arrival rate is $\lambda$,
\item
the service times of the packets $X_{1}$,~$X_{2},\cdots$ are i.i.d. random variables with the first moment $\overline{X}$ and the second moment $\overline{X^2}$, and
\item
the vacation times $V_{1}$,~$V_{2},\cdots$ are i.i.d. random variables with the first moment $\overline{V}$ and the second moment $\overline{V^2}$.
\end{enumerate}

Under the gated-limited service discipline, up to $M$  packets waiting outside the gate will enter the gate at the end of each busy period, and they will be served in the next busy period. After each busy period, the server takes a vacation. When the vacation terminates, the server returns to serve the packets if the buffer inside the gate is not empty; otherwise, the server takes another vacation.

\begin{figure*}[!t]
\centering
\includegraphics[width=0.8\textwidth]{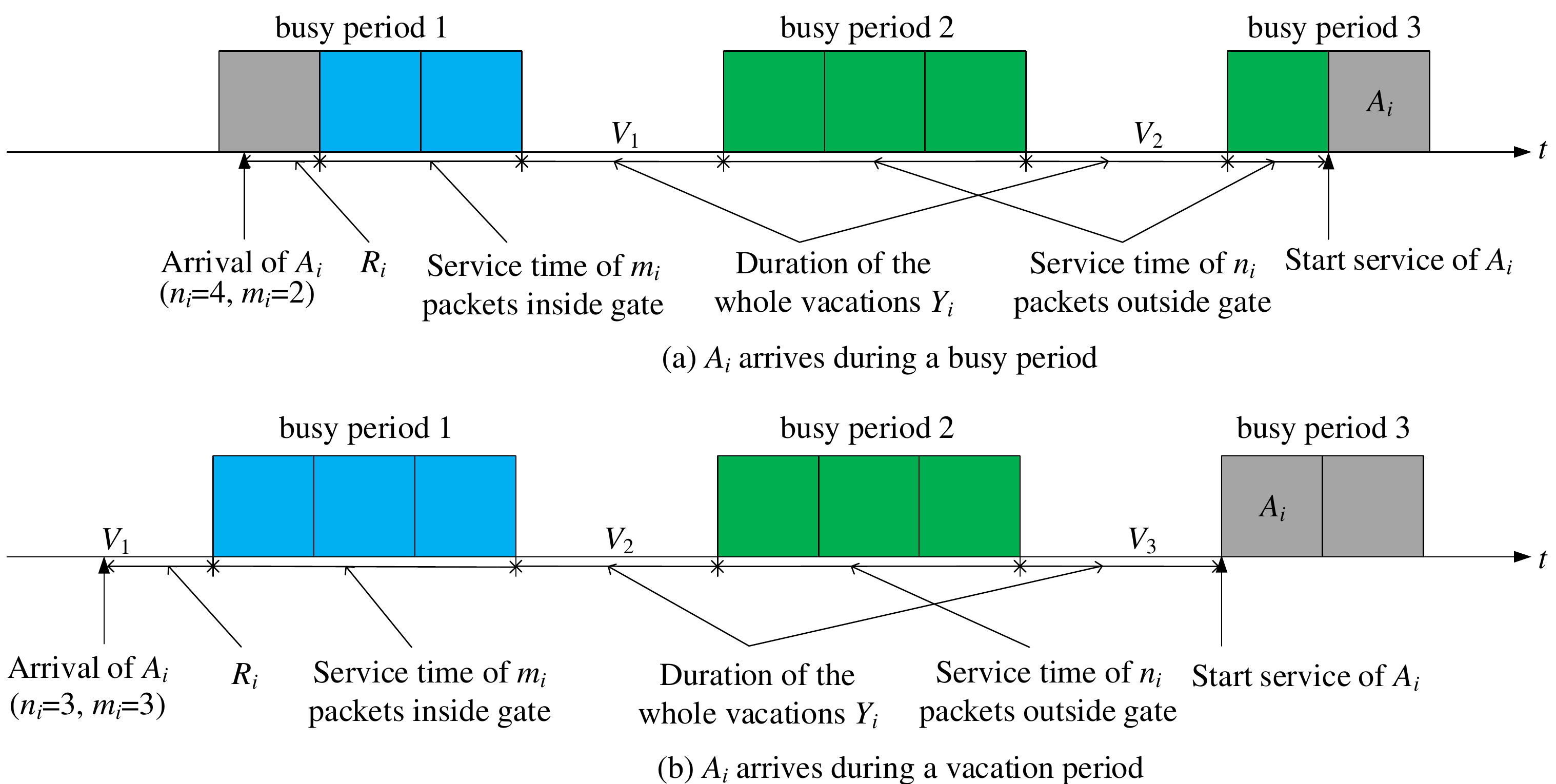}
\caption{Waiting time of the $i$-th packet $A_i$, where $M=3$.}
\label{fig4}
\end{figure*}

A cycle starts at the end of a busy period, and consists of a vacation period followed by another busy period. As Fig.~\ref{fig4} illustrates, the $i$-th packet $A_i$ may arrive at the system during a busy period or a vacation period. The following definitions pertaining to busy periods will be adopted in the derivation of mean waiting time of the packets:
\begin{description}[\IEEEsetlabelwidth{$K=k_2$}]
\item[$B$]
A busy period.
\item[$K=k$]
The number of packets served in a busy period, where $k=0,1,2,\cdots,M$.
\item[$B_k$]
A busy period, during which $k$ packets are served, where $k=0,1,2,\cdots,M$ ($B_{0}$ happens when a vacation finishes while the buffer inside the gate is empty).
\item[$b_k$]
The probability that a busy period is a $B_k$.
\item[$P_k$]
The probability that a packet is served in a $B_k$, where $k=0,1,2,\cdots,M$.
\item[${\Delta}_{i}$]
The number of packets served ahead of the $i$-th packet $A_i$ in the same busy period.
\end{description}

We need the following two lemmas to facilitate the derivation of the mean waiting time of packets.

\newtheorem{myLemma}{Lemma}
\begin{myLemma}
The probability that the $i$-th packet $A_i$ is served in a busy period $B_k$ is given by
\begin{equation}
P_k=\frac{k{b}_{k}}{\overline{K}}
\label{III-A-Lemma1}
\end{equation}
\end{myLemma}
\begin{IEEEproof}
Suppose there are ${\theta}_{k}$ busy periods $B_k$ during a time interval $\left[0,T\right]$, where $k=0,1,2,\cdots,M$. The probability $b_k$ that a busy period is a $B_k$ is defined by
\begin{equation*}
b_k=\lim_{T\to\infty}\frac{{\theta}_{k}}{\sum_{k=0}^M{\theta}_{k}}
\end{equation*}
During the time interval $\left[0,T\right]$, the number of packets that are served in all ${\theta}_{k}$  busy periods $B_k$ is $k{\theta}_{k}$, and the total number of packets served in the interval $\left[0,T\right]$ is $\sum_{k=1}^Mk{\theta}_{k}$. It follows that the probability $P_k$ that the $i$-th packet $A_i$  is served in a busy period $B_k$ can be obtained as follows
\begin{align*}
P_k&=\lim_{T\to\infty}\frac{k{\theta}_{k}}{\sum_{k=1}^M{k{\theta}_{k}}}=\lim_{T\to\infty}\frac{k{\times}\frac{{\theta}_k}{\sum_{k=0}^M{\theta}_k}}{\sum_{k=1}^Mk{\times}\frac{{\theta}_k}{\sum_{k=0}^M{\theta}_k}}\\
   &=\frac{kb_k}{\sum_{k=1}^Mkb_k}=\frac{kb_k}{\overline{K}}.
\end{align*}
\end{IEEEproof}

\newtheorem{myLemma1}[myLemma]{Lemma}
\begin{myLemma1}
The mean number of packets served ahead of the $i$-th packet $A_{i}$ in the same busy period is given by
\begin{equation}
E\left[{\Delta}_{i}\right]=\frac{\overline{K^2}-\overline{K}}{2\overline{K}}.
\label{III-A-Lemma2}
\end{equation}
\end{myLemma1}
\begin{IEEEproof}
Conditioning on the event that packet $A_i$ is served in a busy period $B_k$, we have
\begin{align*}
E\left[{\Delta}_{i}\right] &=\sum_{k=1}^ME\left[{\Delta}_{i}|A_i~is~served~in~a~B_k\right]P_k \\
                           &=\sum_{k=1}^M\frac{0+1+\cdots+(k-1)}{k}\frac{kb_k}{\overline{K}}    \\
                           &=\sum_{k=1}^M\frac{\left(k^2-k\right)b_k}{2\overline{K}}=\frac{\overline{K^2}-\overline{K}}{2\overline{K}}.
\end{align*}
\end{IEEEproof}

As Fig.~\ref{fig4} shows, it may take the server several busy periods to clear all packets waiting in the buffer ahead of packet $A_i$. Since the server can only transmit up to $M$ packets in each busy period, before the starting of service, the waiting time of packet $A_i$ includes the following components:
\begin{enumerate}[{(1)}]
\item
The time to complete current service or current vacation. When the packet $A_i$ arrives, the residual time, either residual service time or residual vacation time, seen by $A_i$ is denoted by $R_i$.
\item
The service times of all $N_i$ packets found waiting in the buffer when $A_i$ arrives.
\item
Besides residual vacation time, the duration of the whole vacation times experienced by $A_i$ before the starting of service is denoted by $Y_i$.
\end{enumerate}
It follows from the similar argument given in \cite{datanetwork}, we have
\begin{equation}
\overline{W}=\overline{R}+N_{Q}\overline{X}+\overline{Y},
\label{mean-waitime}
\end{equation}
and
\begin{equation}
\overline{R}=E\left[R_i\right]=\frac{\lambda\overline{X^2}}{2}+\frac{\left(1-\rho\right)\overline{V^2}}{2\overline{V}},
\label{mean-residual}
\end{equation}
where $N_Q=E\left[N_i\right]=\lambda\overline{W}$ is the mean queue length, and $\rho=\lambda\overline{X}$ is the traffic load. The key to derive the mean waiting time (\ref{mean-waitime}) is the third term $\overline{Y}=E\left[{Y_i}\right]$, which is given in the proof of the following theorem.

\newtheorem{myTheorem}{Theorem}
\begin{myTheorem}
The mean waiting time of an M/G/1 queue with vacations and gated-limited service discipline is given by
\begin{equation}
\overline{W}=\frac{\frac{\lambda\overline{X^2}}{2}+\frac{\left(1-\rho\right)\overline{V^2}}{2\overline{V}}+\left[1-\frac{\left(1+\rho\right)\left(\overline{K^2}-\overline{K}\right)}{2M\overline{K}}-\frac{\lambda\overline{V}}{M}\right]\overline{V}}{1-\rho-\frac{\lambda\overline{V}}{M}}.
\label{III-A-Theorem1}
\end{equation}
\end{myTheorem}
\begin{IEEEproof}
Suppose the system is in state $(n_i,m_i)$ when packet $A_i$ arrives, meaning that the number of packets waiting in the buffer are $n_i$ outside and $m_i$ inside the gate. After $A_i$ arrives, all $m_i$ packets inside the gate are sent out during the first busy period, at the end of which the first $M$ of the $n_i$ packets enters the gate. The packet $A_i$ enters the gate at the end of the $\left(\lfloor{\left.{n}_{i}\middle/M\right.}\rfloor+1\right)$-th busy period, and is sent out during the  $\left(\lfloor{\left.{n}_{i}\middle/M\right.}\rfloor+2\right)$-th busy period. That is, the number of whole vacations that $A_i$ has to experience before the starting of service is  $\lfloor{\left.{n}_{i}\middle/M\right.}\rfloor+1$, where $\lfloor{x}\rfloor$ is the largest integer smaller than $x$.

For example, as Fig.~\ref{fig4}(a) shows, the state of system is $(n_i,m_i)=(4,2)$ upon the arrival of $A_i$ and $M$ is three, thus $A_i$ has to wait for $\lfloor{\left.{n}_{i}\middle/M\right.}\rfloor+1=\lfloor{4/3}\rfloor+1=2$ whole vacation times in the buffer before it can be transmitted. It's the same as that in Fig.~\ref{fig4}(b). Thus, we have
\begin{equation}
\overline{Y}=\left(1+E\left[ \left\lfloor{\frac{n_i}{M}}\right\rfloor \right]\right)\overline{V}.
\label{total-vacation}
\end{equation}
In the $\left(\lfloor{\left.{n}_{i}\middle/M\right.}\rfloor+2\right)$-th busy period, the number of packets transmitted ahead of $A_i$ is given by ${\Delta}_{i}=n_i-{\lfloor}{\frac{n_i}{M}}{\rfloor}M$.
For example, as Fig.~\ref{fig4}(a) shows, packet $A_i$ is the second packet served in the third busy period. Since $n_{i}=4$ upon the arrival of $A_i$ and $M=3$, it follows that ${\Delta}_{i}=n_i-{\lfloor}{\frac{n_i}{M}}{\rfloor}M=4-3=1$. However, in Fig.~\ref{fig4}(b), there is no packet transmitted before $A_i$ in the third busy period since the system state is $(n_i,m_i)=(3,3)$ when $A_i$ arrives, and in this case ${\Delta}_{i}=n_i-{\lfloor}{\frac{n_i}{M}}{\rfloor}M=3-3=0$. Therefore, by definition, we have
\begin{equation}
E\left[{\left\lfloor\frac{n_i}{M}\right\rfloor}\right]=\frac{E\left[{n_i}\right]-E\left[{{\Delta}_{i}}\right]}{M}.
\label{floor-mean}
\end{equation}
Notice that the mean queue length $N_Q$ is the sum of the mean number of packets waiting outside the gate $\overline{n}=E[n_i]$ and that waiting inside the gate $\overline{m}=E[m_i]$. It follows that
\begin{equation}
\overline{n}=E[n_i]=N_Q-\overline{m}.
\label{meanql-outside}
\end{equation}
Since the packet $A_i$ is moved into the gate at the end of the $\left(\lfloor{\left.{n}_{i}\middle/M\right.}\rfloor+1\right)$-th busy period and served in the $\left(\lfloor{\left.{n}_{i}\middle/M\right.}\rfloor+2\right)$-th busy period, the waiting time of $A_i$ inside the gate, denoted as $W_{in}^i$, includes the vacation time $V$ between these two busy periods, and the total service time of ${\Delta}_{i}$ packets transmitted ahead of $A_i$ in the $\left(\lfloor{\left.{n}_{i}\middle/M\right.}\rfloor+2\right)$-th busy period. Thus, the mean waiting time of $A_i$ inside the gate is given by
\begin{equation*}
\overline{W}_{in}=E\left[{W_{in}^i}\right]=\overline{V}+E\left[{{\Delta}_i}\right]\overline{X}.
\end{equation*}
From Little's Law and Lemma 2, we obtain the following mean number of packets waiting inside the gate:
\begin{equation}
\overline{m}=\lambda\overline{W}_{in}=\lambda\overline{V}+\frac{\rho\left({\overline{K^2}-\overline{K}}\right)}{2\overline{K}}.
\label{meanql-inside}
\end{equation}
The theorem is established by combining (\ref{mean-waitime})-(\ref{mean-residual}) and (\ref{total-vacation})-(\ref{meanql-inside}).
\end{IEEEproof}

Suppose the distribution of service time $X$ is given. The evaluation of the mean waiting time (\ref{III-A-Theorem1}) requires the first two moments of the vacation time $V$ and the number of packets $K$ transmitted in a busy period. Intuitively, they are related to each other because the random variable $K$ is dependent on the number of arrivals during the vacation time $V$. Focusing on the application of the above theorem to EPONs, we will discuss the relationship between the first two moments of $V$ and $K$ in the next Section \uppercase\expandafter{\romannumeral3}-B.

\subsection{Mean Packet Waiting Time of EPONs with Gated-Limited Service Discipline}
In this subsection, we apply the result of Theorem 1 to calculate the mean packet waiting time of an ONU in the gated-limited service EPON, where the rate of the traffic input to the network is ${\lambda}_E$ and to each ONU is $\lambda=\left.{\lambda}_{E}\middle/N\right.$. We assume that the distribution of the packet transmission time $X$ is given.

1) \emph{Moments of Vacation Time V of An ONU:} An ONU is busy with probability $\rho=\lambda\overline{X}$ and idle with probability $1-\rho$. The mean busy period of an ONU is given by
\begin{equation}
\overline{B}=E[B]=\frac{\rho\overline{V}}{1-\rho}.
\label{mean-busyperiod-1}
\end{equation}
In an EPON with $N$ ONUs, the vacation time of an ONU is equal to the TWs of other $(N-1)$ ONUs plus $NG$. According to our assumption A2, the TWs are i.i.d. random variables. By definition, we have
\begin{equation*}
\overline{V}=(N-1)\overline{B}+NG=(N-1)\frac{\frac{{\rho}_E}{N}\overline{V}}{1-\frac{{\rho}_E}{N}}+NG,
\end{equation*}
where ${\rho}_E={\lambda}_E\overline{X}=N\rho$ is the offered load to the EPON. After some reconfigurations, the first moment of the vacation time $V$ for an ONU is given by
\begin{equation}
\overline{V}=\frac{N-{\rho}_E}{1-{\rho}_E}G.
\label{mean-ONU-vacation}
\end{equation}
Similarly, the second moment of the vacation time $V$ for an ONU is defined as follows
\begin{equation}
\overline{V^2}={\overline{V}}^2+{\sigma}_V^2={\overline{V}}^2+(N-1){\sigma}_B^2,
\label{secmoment-ONUvacation-1}
\end{equation}
where ${\sigma}_V^2$ and ${\sigma}_B^2$ are the variances of $V$ and $B$, respectively. Recall that $B_k$ is a busy period during which $k$ packets are transmitted. It follows that $B_k=\sum_{i=1}^k{X_i}$, in which $X_1,X_2,\cdots,X_k$ are i.i.d. random variables. Let $X^*(\theta)$ and $B^*(\theta)$ be the Laplace-Stieltjes transforms of the probability density function (PDF) of the service time $X$ and the busy period $B$, respectively. They are related as follows:
\begin{align*}
B^*(\theta) &=E\left[{e^{-{\theta}B}}\right]=E\left[{E\left[{e^{-{\theta}B}|B_k}\right]}\right]=\sum_{k=0}^ME\left[{e^{-{\theta}B_k}}\right]b_k  \\
            &=\sum_{k=0}^M\left({\prod_{i=1}^kE\left[{e^{-{\theta}X_i}}\right]}\right)b_k=\sum_{k=0}^M{\left[{X^*(\theta)}\right]}^k{b_k}  \\
            &=F\left[{X^*(\theta)}\right],
\end{align*}
where $F(z)=\sum_{k=0}^M{b_k}z^k$ is the generating function of $b_k$. Therefore, the variance of the busy period ${\sigma}_B^2$ can be obtained by
\begin{equation}
\begin{split}
{\sigma}_B^2 &=B^{*\prime\prime}(0)-{\left[-B^{*\prime}(0)\right]}^2  \\
             &=F^{\prime\prime}(1){\left[X^{*\prime}(0)\right]}^2+F^{\prime}(1)X^{*\prime\prime}(0)-{\left[F^{\prime}(1)X^{*\prime}(0)\right]}^2  \\
             &=\left({\overline{K^2}-\overline{K}}\right){\overline{X}}^2+\overline{K}\overline{X^2}-{\left({\overline{K}\overline{X}}\right)}^2  \\
             &={\overline{X}}^2\left({\overline{K^2}-{\overline{K}}^2}\right)+\overline{K}\left({\overline{X^2}-{\overline{X}}^2}\right).
\label{busyperiod-var}
\end{split}
\end{equation}
Substituting (\ref{busyperiod-var}) into (\ref{secmoment-ONUvacation-1}), we obtain the following expression of $\overline{V^2}$:
\begin{equation}
\overline{V^2}={\overline{V}}^2+(N-1)\left[{{\overline{X}}^2\left({\overline{K^2}-\overline{K}}\right)+\overline{K}\left({\overline{X^2}-\overline{X}}\right)}\right].
\label{secmoment-ONUvacation-2}
\end{equation}
From (\ref{mean-ONU-vacation}) and (\ref{secmoment-ONUvacation-2}), we know from (\ref{III-A-Theorem1}) that the mean packet waiting time of an ONU can now be determined by the first two moments $\overline{K}$ and $\overline{K^2}$ of the number of packets transmitted in a busy period.

\begin{figure}[!t]
\centering
\includegraphics[width=0.45\textwidth]{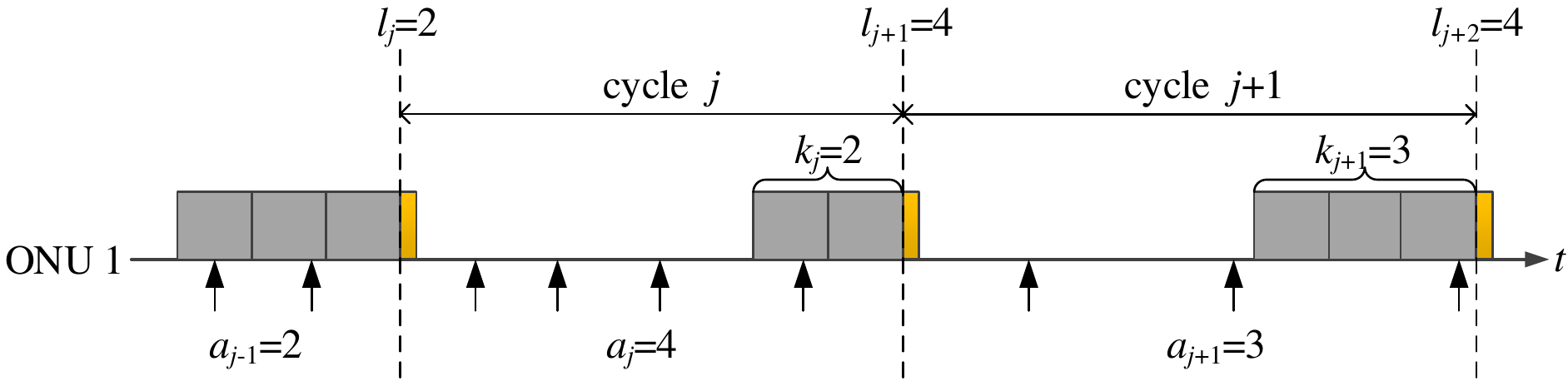}
\caption{Upstream transmission process of an ONU, where $M=3$.}
\label{fig5}
\end{figure}

2) \emph{Moments of Number of Packets K Transmitted in A Busy Period:} The first moment of $K$ can be easily derived from (\ref{mean-busyperiod-1}) and (\ref{mean-ONU-vacation}), and given as follows:
\begin{equation}
\overline{K}=\frac{\overline{B}}{\overline{X}}=\frac{\lambda\overline{V}}{1-\rho}=\frac{\frac{{\lambda}_E}{N}\overline{V}}{1-\frac{{\rho}_E}{N}}=\frac{{\lambda}_{E}G}{1-{\rho}_E}.
\label{mean-K}
\end{equation}
The derivation of the second moment $\overline{K^2}$, however, has to resort to the discrete time Markov chain embedded in the epochs at the end of busy periods. As Fig.~\ref{fig5} shows, the upstream transmission process of an ONU is a sequence of cycles. For example, at the end of cycle $j-1$ and the beginning of cycle $j$, the ONU reports its queue length, denoted as $l_j$, to the OLT. According to this report, the OLT determines the size of the TW in cycle $j$, denoted as $k_j$, as follows: $k_j=M$ if ${l_j}\geq{M}$, and $k_j=l_j$, if $l_j<M$.
That is, the TW size $k_j$ in cycle $j$ is determined by the queue length $l_j$ at the start of cycle $j$ and given as follows
\begin{equation}
k_j=l_j-{\left({l_j-M}\right)}^{+},
\label{kj-lj}
\end{equation}
where ${\left(l_{j}-M\right)}^{+}{\triangleq}\max{\left\{l_{j}-M,0\right\}}$ is the number of reported packets that are not transmitted in cycle $j$.

Let $q_{n}=\lim_{j\to\infty}Pr\{l_{j}=n\}$. Recall that $b_k$ is the probability that $k$ packets are served in a busy period. According to (\ref{kj-lj}), we have
\begin{equation}
b_k=\left\{
\begin{array}{cl}
q_k, &                  k=0,1,\cdots,M-1 \\
1-\sum_{k=0}^{M-1}q_k,& k=M
\end{array}
\right.
\label{bk-qk}
\end{equation}
Thus, the second moment of $K$ can be obtained based on the distribution of the queue length $q_n$ as follows
\begin{equation}
\overline{K^2}=\sum_{k=0}^Mk^2b_k=\sum_{k=0}^{M-1}k^2q_k+M^2\left({1-\sum_{k=0}^{M-1}q_k}\right).
\label{K2-qk}
\end{equation}
On the other hand, the queue length $l_{j+1}$  at the start point of cycle $j+1$ is determined by the number of packets $k_j$ transmitted during the busy period of cycle $j$ and the number of arrivals $a_j$ during cycle $j$. Thus, from (\ref{kj-lj}), the queue length at the start point of each cycle satisfies the following Lindley's equation:
\begin{equation}
l_{j+1}=l_j-k_j+a_j={\left({l_j-M}\right)}^{+}+a_j.
\label{ql-relation}
\end{equation}

Let $h_n=\lim_{j\to\infty}Pr\{a_j=n\}$ be the probability that there are $n$ arrivals during a cycle time $C$. We immediately derive the following equilibrium equation from (\ref{ql-relation}):
\begin{equation*}
q_n=\sum_{i=0}^{M-1}q_ih_n+\sum_{i=M}^{M+n}q_ih_{n+M-i},
\end{equation*}
from which we obtain the generating function of queue length:
\begin{equation}
Q(z){\triangleq}\sum_{n=0}^{\infty}q_nz^n=\frac{\left[{\sum_{i=0}^{M-1}q_i\left({z^M-z^i}\right)}\right]H\left(z\right)}{z^M-H(z)},
\label{genfun-qn}
\end{equation}
where $H\left(z\right){\triangleq}\sum_{n=0}^{\infty}h_{n}z^{n}$ is the generating function of $h_n$.

According to our assumption A3 that the packet arrival process of each ONU is a Poisson process, the distribution $h_n$ is completely determined by the cycle time distribution. Let $c(t)$ be the PDF of the cycle time $C$. We have
\begin{equation}
H(z)=\sum_{n=0}^{\infty}\left[{\int_{0}^{\infty}\frac{{\left({{\lambda}t}\right)}^n}{n!}e^{-{\lambda}t}c\left(t\right)dt}\right]z^n=C^{*}\left[{{\lambda}\left(1-z\right)}\right].
\label{genfun-hn}
\end{equation}

A cycle consists of a vacation period and a busy period, which means that the cycle time is the sum of $NG$ and the duration of the TWs of $N$ ONUs, which are i.i.d. random variables according to our assumption A2. It follows that the distribution of cycle time is approximately a Gaussian distribution according to the central limit theorem \cite{central-limit}. The Laplace-Stieltjes transform of the cycle time distribution is given by
\begin{equation}
C^{*}(\theta)=exp\left[{-{\mu}_{C}\theta+\frac{1}{2}{\sigma}_{C}^2{\theta}^2}\right],
\label{LST-cycle}
\end{equation}
where the mean cycle time can be obtained from (\ref{mean-ONU-vacation}), and is given as follows:
\begin{equation}
{\mu}_{C}=\frac{\overline{V}}{1-\rho}=\frac{\frac{\left(N-{\rho}_E\right)G}{1-{\rho}_E}}{1-\frac{{\rho}_E}{N}}=\frac{NG}{1-{\rho}_{E}},
\label{mean-cycle}
\end{equation}
while the variance of the cycle time $C$ is determined by the variance of busy period (\ref{busyperiod-var}):
\begin{equation}
{\sigma}_C^2=N{\sigma}_B^2=N\left[{{\overline{X}}^2\left({\overline{K^2}-{\overline{K}}^2}\right)+\overline{K}\left({\overline{X^2}-{\overline{X}}^2}\right)}\right].
\label{var-cycle}
\end{equation}
We know that the second moment $\overline{K^2}$ in (\ref{K2-qk}) is coupled to the queue length probability $q_n$, for $n=0,1,\cdots,M-1$. For a given $\overline{K^2}$, according to Rouche's theorem and Lagrange's theorem \cite{TakagiQueueingBook}, we can first solve $q_n~(n=0,1,\cdots,M-1)$ from (\ref{genfun-qn})-(\ref{var-cycle}), and then update $\overline{K^2}$ by substituting $q_n$ into (\ref{K2-qk}). Repeatedly applying this iterative procedure, we can obtain the value of $\overline{K^2}$ and then obtain the mean waiting time (\ref{III-A-Theorem1}) of an ONU by combining (\ref{mean-ONU-vacation}), (\ref{secmoment-ONUvacation-2}), (\ref{mean-K}), and $\overline{K^2}$. The procedure that numerically calculates $\overline{K^2}$ and the mean waiting time is given in APPENDIX A. In the next section, we seek a systematic rule to select the optimum TW size for each ONU of the EPON that satisfies practical operational requirements of EPONs with gated-limited service.

\section{Optimum Transmission Window Size}
As we mentioned in Section I, EPON users usually have to sign a SLA with the network operator to specify the upstream traffic rate. Suppose all ONUs are statistically identical, and each ONU subscribes to a SLA with a maximum traffic rate $\left.{{\lambda}_E^{*}}\middle/{N}\right.$, where ${\lambda}_E^{*}$ is the total subscribed traffic rate of all the ONUs, which is less than $\left.1\middle/\overline{X}\right.$ . An ONU is a disciplined user if its input traffic rate is in the admissible region $\lambda={\lambda}_E/N\in\left[{0,{\lambda}_E^{*}/N}\right]$. Otherwise, it is considered a malicious user. An EPON system is \emph{regular} if all the ONUs are disciplined users. In this section, we first describe the methodology and procedure to select an optimum TW size $M$ for a given traffic rate ${\lambda}_E^{*}/N$. We then discuss the stability and delay performance of EPON system under this selected  TW size $M$.

As we mentioned in Section \uppercase\expandafter{\romannumeral2}, the purpose of limiting the TW size $M$ is twofold: to guarantee that the mean delays experienced by disciplined users are bounded, and to penalize malicious users. With gated-limited service discipline, the TW size $M$ limits the maximum number of packets that can be served in a busy period. Ideally, a proper TW size $M$ ensures that all packets arrived at a disciplined ONU during a cycle time can be completely served in the next busy period. To achieve this goal, the probability that the queue length at the beginning of a cycle exceeds the limit $M$ should be kept very small. Thus, the criterion for the selection of TW size $M$ is given by
\begin{equation}
Pr\{l{\geq}M\}=\lim_{j\to\infty}Pr\{l_j{\geq}M\}{\leq}\varepsilon,
\label{IV-cretirion-org}
\end{equation}
for some positive $\varepsilon\ll1$, when $\lambda={\lambda}_{E}/N\in\left[0,{\lambda}_E^{*}/N\right]$. In the following, we show that an optimum TW size $M$ that satisfies the criterion (\ref{IV-cretirion-org}) can be selected by using the Chernoff bound of queue length.

\subsection{Chernoff Bound of Queue Length}
The Chernoff bound of the tail distribution of queue length $l$ at the beginning of a cycle is given as follows \cite{chernoffbound}:
\begin{equation}
Pr\{l{\geq}{\mu}_{l}+t\}=Pr\{z^l{\geq}z^{{\mu}_{l}+t}\}{\leq}\frac{E\left[z^l\right]}{z^{{\mu}_{l}+t}},
\label{IV-A-chernoff}
\end{equation}
for any $z>1$, where $E[z^l]=Q(z)$ is the generating function of the queue length distribution and ${\mu}_l=E[l]$ is the mean queue length.

Suppose all the ONUs are disciplined users with input traffic rate $\lambda={\lambda}_E/N\in\left[{0,{\lambda}_E^{*}/N}\right]$, and the TW size $M$ satisfies the criterion (\ref{IV-cretirion-org}), then each time the queue length reported by an ONU should be typically smaller than $M$ with a high probability $1-\varepsilon$. It follows that equations (\ref{kj-lj}) and (\ref{ql-relation}) will respectively degenerate to the following approximate equations:
\begin{align}
l_j&{\approx}k_j,    \label{app-lj-kj} \\
l_{j+1}&{\approx}a_j,
\label{app-lj-aj}
\end{align}
which implies that the following generating functions of $l_j$, $k_j$ and $a_j$ are approximately equal:
\begin{equation}
Q(z){\approx}F(z){\approx}H(z).
\label{app-genfun-qnbnhn}
\end{equation}
Thus, according to (\ref{genfun-hn})-(\ref{var-cycle}), we have
\begin{align}
Q(z){\approx}H(z)=&exp\left[{-{\lambda}{\mu}_{C}(1-z)+\frac{1}{2}{\lambda}^2{\sigma}_C^2(1-z)^2}\right]  \nonumber\\
     =&exp\left\{-\frac{{\lambda}_{E}G\left(1-z\right)}{1-{\rho}_{E}}+\left[{\overline{X}}^2\big({\overline{K^2}-{\overline{K}}^2}\big)\right.\right. \nonumber\\
                &\left.\left.+\overline{K}\big({\overline{X^2}-{\overline{X}}^2}\big)\right]\frac{{\lambda}_{E}^2(1-z)^2}{2N}\right\}.
\label{Qz-expre-1}
\end{align}
In this equation, the second moment of the number of packets served in each busy period can be obtained by
\begin{equation}
\overline{K^2}=F^{\prime\prime}(1)+F^{\prime}(1){\approx}H^{\prime\prime}(1)+H^{\prime}(1).
\label{K2-expre-1}
\end{equation}
Substituting (\ref{Qz-expre-1}) into (\ref{K2-expre-1}), we have
\begin{align*}
\overline{K^2}=&{\left({\frac{{\lambda}_{E}G}{1-{\rho}_{E}}}\right)}^2+\frac{{\rho}_E^2}{N}\left[{\overline{K^2}-{\left({\frac{{\lambda}_{E}G}{1-{\rho}_{E}}}\right)}^2}\right] \\
               &+\frac{{\lambda}_{E}^{3}G}{N\left({1-{\rho}_E}\right)}\left({\overline{X^2}-{\overline{X}}^2}\right)+\frac{{\lambda}_{E}G}{1-{\rho}_{E}},
\end{align*}
which yields
\begin{equation}
\overline{K^2}={\left({\frac{{\lambda}_{E}G}{1-{\rho}_{E}}}\right)}^2+\frac{\frac{{\lambda}_{E}^{3}G}{N\left({1-{\rho}_E}\right)}\left({\overline{X^2}-{\overline{X}}^2}\right)+\frac{{\lambda}_{E}G}{1-{\rho}_{E}}}{1-\frac{{\rho}_E^2}{N}}.
\label{K2-expre-appro}
\end{equation}
Substituting (\ref{K2-expre-appro}) into (\ref{Qz-expre-1}), we obtain $Q(z)$ in the \emph{regular} case as follows:
\begin{align}
&Q(z){\approx}  \nonumber \\
&exp\left[-\frac{{\lambda}_{E}G}{1-{\rho}_{E}}(1-z)+\frac{{\lambda}_{E}^{3}G\overline{X^2}}{2\left({1-{\rho}_E}\right)\left({N-{\rho}_E^2}\right)}{(1-z)}^2\right].
\label{Qz-expre-appro}
\end{align}
The mean and variance of queue length $l$ are respectively given as follows:
\begin{equation}
{\mu}_l=Q^{\prime}\left(1\right)={\lambda}{\mu}_{C}=\frac{{\lambda}_{E}G}{1-{\rho}_{E}},
\label{mean-ql}
\end{equation}
and
\begin{align}
{\sigma}_l^2 &=Q^{\prime\prime}(1)+Q^{\prime}(1)-{\left[{Q^{\prime}(1)}\right]}^2={\lambda}^2{\sigma}_C^2+{\lambda}{\mu}_C   \nonumber\\
             &=\frac{{\lambda}_{E}^{3}G\overline{X^2}}{\left({1-{\rho}_E}\right)\left({N-{\rho}_E^2}\right)}+\frac{{\lambda}_{E}G}{1-{\rho}_{E}}.
\label{var-ql}
\end{align}
It follows from the first equation of generating function of queue length in (\ref{Qz-expre-1}), the Chernoff bound (\ref{IV-A-chernoff}) is given by
\begin{align*}
&Pr\{l{\geq}{\mu}_l+t\}{\leq}  \\
&exp\left[{-\left({{\mu}_l+t}\right){\log}z+{\lambda}{\mu}_{C}\left(z-1\right)+\frac{1}{2}{\lambda}^2{\sigma}_C^2{(z-1)}^2}\right],
\end{align*}
for any $z>1$. Substituting $t=M-{\mu}_l$ into the above Chernoff bound, the criterion (\ref{IV-cretirion-org}) can be fulfilled if $M$ is the smallest integer that satisfies the following inequality:
\begin{align}
&Pr\{l{\geq}M\}{\leq}    \nonumber \\
&\inf_{z>1}\left\{exp\left[{-M{\log}z+{\lambda}{\mu}_C\left(z-1\right)+\frac{1}{2}{\lambda}^2{\sigma}_C^2{\left(z-1\right)}^2}\right]\right\}\!{\leq}{\varepsilon},
\label{IV-A-Chernoff-2}
\end{align}
where ${\lambda}{\in}\left[{0,{\lambda}_E^{*}/N}\right]$. We discuss the procedure to find the optimum TW size $M^{*}$ that satisfies (\ref{IV-A-Chernoff-2}) in the next subsection.

\subsection{Optimum Transmission Window Size}
Solving the optimum TW size $M^{*}$ from (\ref{IV-A-Chernoff-2}) involves a complicated transcendental equation, therefore it can only be solved numerically. To initialize the computation procedure, we provide a lower bound $M_1$ and an upper bound $M_2$ of $M^{*}$ in the following theorem.
\newtheorem{myTheorem1}[myTheorem]{Theorem}
\begin{myTheorem1}
The optimum TW size $M^{*}$ that satisfies (\ref{IV-A-Chernoff-2}) is bounded by
\begin{equation}
\left\lceil{{\mu}_l+\!\lambda{\sigma}_C\sqrt{2\alpha}}\right\rceil\!=\!M_1{\leq}M^{*}{\leq}M_2\!=\!\left\lceil{{\mu}_l+\alpha+\!\sqrt{{\alpha}^2+2\alpha{\sigma}_l^2}}\right\rceil
\label{IV-B-theorem2}
\end{equation}
where $\alpha=\log{{\varepsilon}^{-1}}$ and $\lambda={\lambda}_E/N{\in}\left[{0,{\lambda}_E^*/N}\right]$.
\end{myTheorem1}
\hfill$\blacksquare$

\begin{figure}[!t]
\centering
\includegraphics[width=0.45\textwidth]{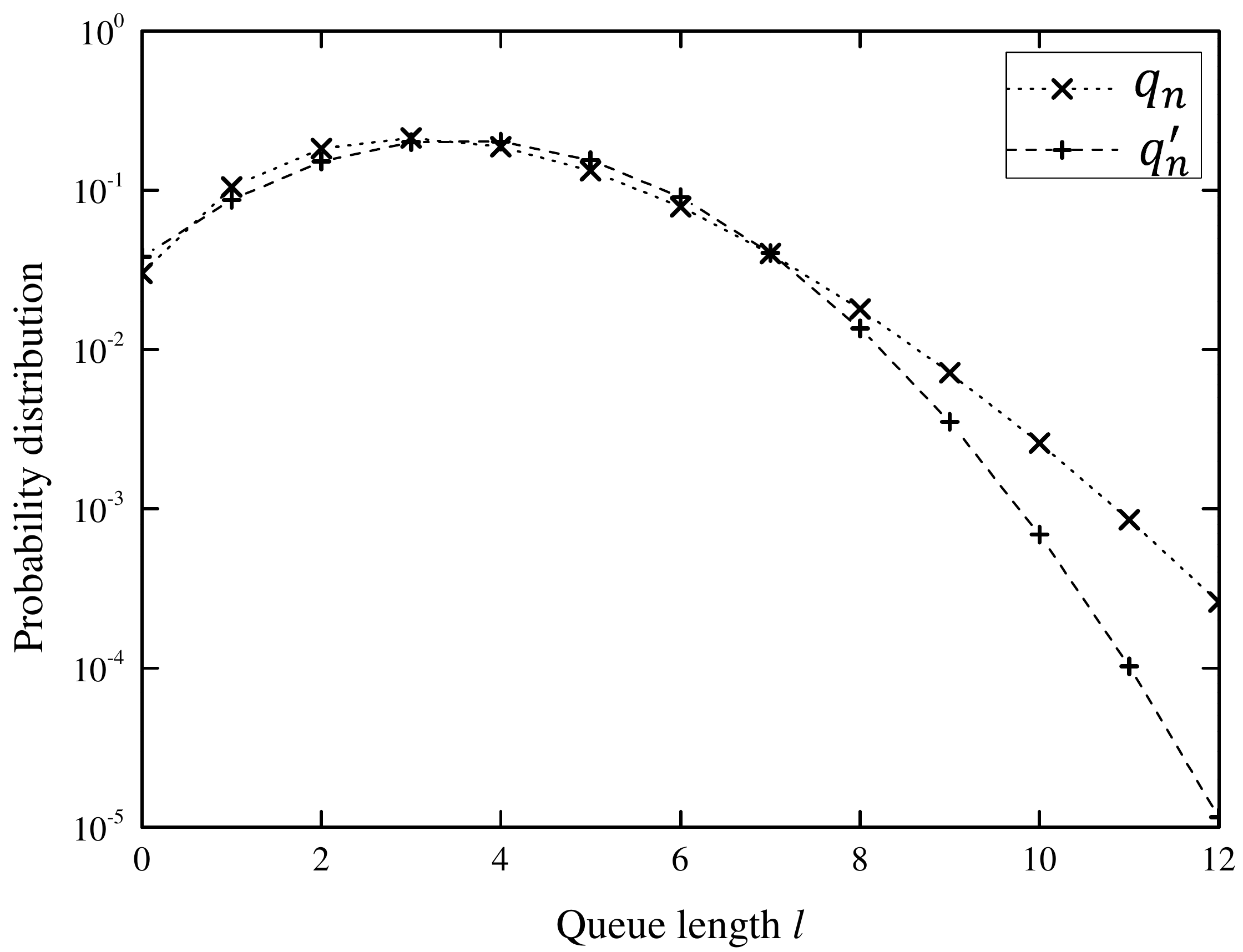}
\caption{Probability distribution $q_n$ and $q_n^{\prime}$ with $\lambda=21.875$ packets/ms.}
\label{fig6}
\end{figure}

The proof of the above theorem is given in APPENDIX B. An accurate approximation of the optimum TW size $M^*$ can be derived from the upper deviation inequality of normal random variables. We know that the cycle time $C$ approaches a normal random variable $\mathcal{N}(\mu_C,\sigma_C^2)$ when $N$ is large. The relation (\ref{app-lj-aj}) indicates that the queue length at the beginning of each cycle is approximately equal to the number of arrivals during a cycle time $C$, or $l\sim{\lambda}C$. As expected, the mean queue length $\mu_l$ given by (\ref{mean-ql}) is the product of the arrival rate $\lambda$ and the mean cycle time $\mu_C$. It is also interesting to note that the variance of the queue length $\sigma_l^2$  given by (\ref{var-ql}) is the sum of $\lambda^2\sigma_C^2$ and the variance of a Poisson random variable with parameter $\mu_l$. Thus, the queue length $l$ can be approximated by a normal random variable $\mathcal{N}\left(\mu_l,\sigma_l^2\right)$ that is ¡°discretized¡± by a Poisson process with rate $\lambda$. That is, we adopt the following approximation of queue length distribution:
\begin{equation}
q_n{\cong}q_n^{\prime}=\frac{1}{\sqrt{2\pi}\sigma_l}\int_{n-\frac{1}{2}}^{n+\frac{1}{2}}e^{-\frac{{\left(x-\mu_l\right)}^2}{2\sigma_l^2}}dx.
\label{appro-qn}
\end{equation}
As Fig.~\ref{fig6} shows, the bigger the gap between these two distributions, the smaller the probability $q_n$, where $q_n$ is obtained through the inverse transform of $Q(z)$ in (\ref{Qz-expre-appro}). We use the same set of parameters, including the number of ONUs $N=32$, a guard time and a REPORT message transmission time $G=1.512{\mu}s$, the first and second moments of service time $\overline{X}=1{\mu}s,\overline{X^2}=1{{\mu}s}^2$, in all figures of this subsection.

It is well-known that any normal random variable $X{\sim}\mathcal{N}(\mu,\sigma^2)$ satisfies the following upper deviation inequality \cite{GaussianTail}:
\begin{equation*}
Pr\{X{\geq}\mu+t\}{\leq}exp\left[{-\frac{t^2}{2{\sigma}^2}}\right],
\end{equation*}
for $t{\geq}0$. Since the distribution of queue length $l$ is close to that of the normal random variable $\mathcal{N}\left(\mu_l,\sigma_l^2\right)$, from the above inequality, the optimum TW size $M^*$ can be estimated by the smallest integer $\hat{M}$ that satisfies the following relation:
\begin{equation*}
Pr\{{l{\geq}\hat{M}}\}{\leq}exp\left[{-\frac{{\left(\hat{M}-\mu_l\right)}^2}{2\sigma_l^2}}\right]{\leq}\varepsilon,
\end{equation*}
and it is explicitly given by
\begin{equation}
\hat{M}=\left\lceil{{\mu}_l+\sigma_l\sqrt{2\alpha}}\right\rceil.
\label{appro-TW}
\end{equation}
The following inequality immediately follows from (\ref{var-ql}) and (\ref{appro-TW}):
\begin{equation*}
\left\lceil{{\mu}_l+\lambda{\sigma}_C\sqrt{2\alpha}}\right\rceil{\leq}\hat{M}{\leq}\left\lceil{{\mu}_l+\alpha+\sqrt{{\alpha}^2+2\alpha{\sigma}_l^2}}\right\rceil.
\end{equation*}
That is, the approximation $\hat{M}$ of the optimum TW window size $M^{*}$ also lies between the two bounds $M_1$ and $M_2$.

As we mentioned before, the optimum TW size $M^*$ that satisfies the inequality (\ref{IV-A-Chernoff-2}) can only be solved numerically from the following equation:
\begin{align*}
&f\left(t,z^{*}\right)=  \nonumber \\
&exp\left[\!-\!\left(\mu_l+t\right)\log{z^*}\!+\!\lambda\mu_C\left(z^*-1\right)\!+\!\frac{1}{2}{\lambda}^2\sigma_C^2{\left(z^*-1\right)}^2\right]\!=\!\varepsilon,
\end{align*}
where $z^*$ is obtained from the proof of theorem 2 in APPENDIX B and given as follows:
\begin{equation}
z^*\!=\!\frac{\sqrt{{\left({\lambda\mu_C-{\lambda}^2\sigma_C^2}\right)}^2+4\left(\mu_l+t\right){\lambda}^2\sigma_C^2}-\left({\lambda\mu_C-{\lambda}^2\sigma_C^2}\right)}{2{\lambda}^2\sigma_C^2}.
\label{optimum-z}
\end{equation}

The following procedure is used to solve the optimum TW size $M^*$ that satisfies the inequality (\ref{IV-A-Chernoff-2}).
\begin{enumerate}[{Step 1:}]
\item
$\lambda=\lambda_E^*/N,M=\hat{M},low=M_1,up=M_2$;
\item
$t=M-\mu_l$, calculate $z^*$ by (\ref{optimum-z});
\item
If $f\left(t,z^*\right)>\varepsilon$, $low=M$; else $up=M$;  \\
/* If $f(t,z^*)$ is too large, we update the lower bound of searching region to decrease $f(t,z^*)$, otherwise we update the upper bound. */
\item
If $\left\lceil{low}\right\rceil{<}\left\lceil{up}\right\rceil$, $M=\left(low+up\right)/2$, go to Step 2;
\item
$M^*=\left\lceil{low}\right\rceil=\left\lceil{up}\right\rceil$, output $M^*$.
\end{enumerate}

In the practical operation of EPON, the parameter $\varepsilon$ can be selected from the region $[0.001,0.1]$, which implies that the buffered packets of an ONU can be emptied with a probability $1-\varepsilon$ between 0.9 to 0.999 at the end of every busy period. A too large $\varepsilon$ causes a too small $M$ that impairs the delay performance of disciplined users. On the other hand, a too small $\varepsilon$ causes a too large $M$ that cannot effectively suppress the capture effect.

\begin{figure}[!t]
\centering
\subfigure[]{
\label{fig7-a}
\includegraphics[width=0.45\textwidth]{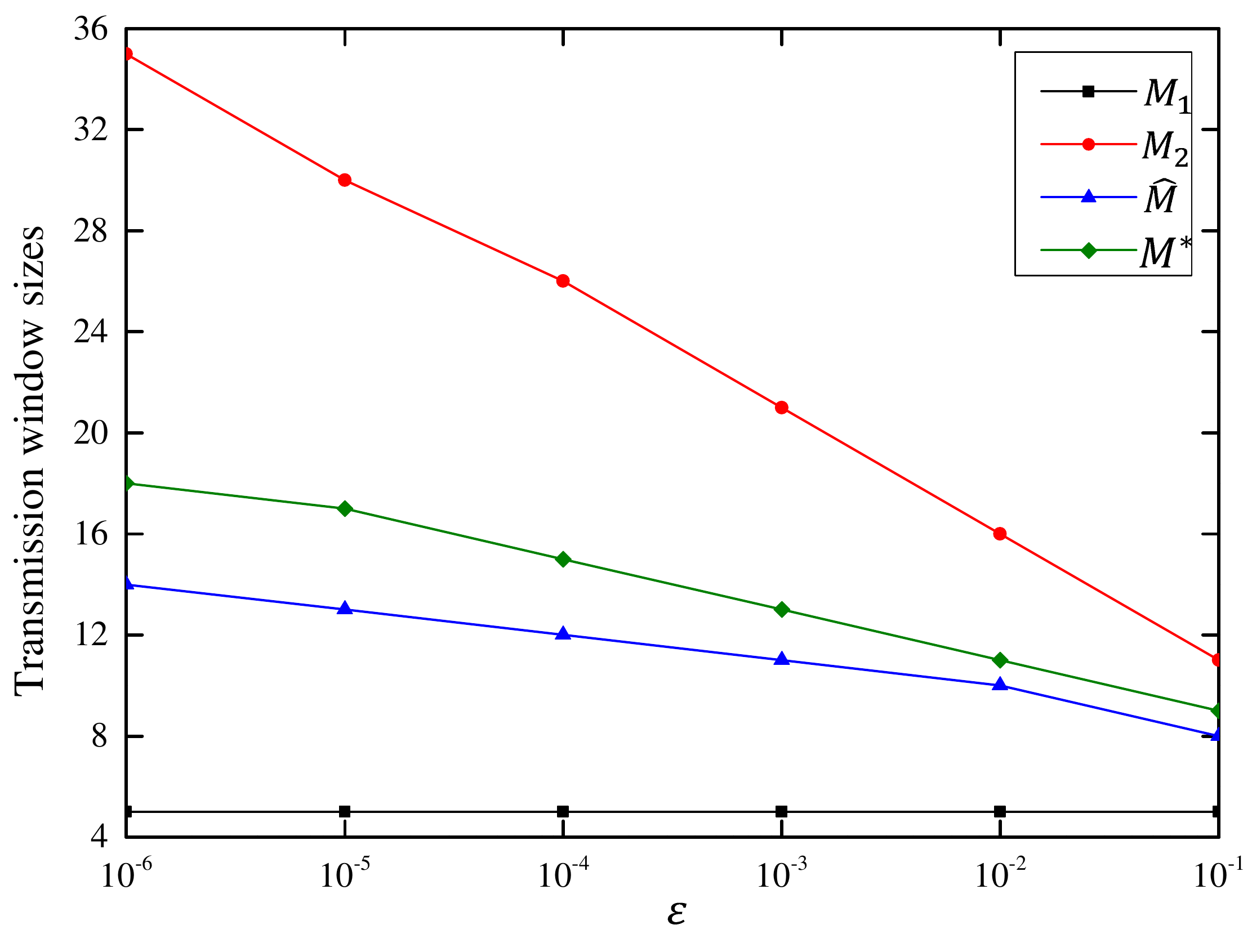}}
\subfigure[]{
\label{fig7-b}
\includegraphics[width=0.45\textwidth]{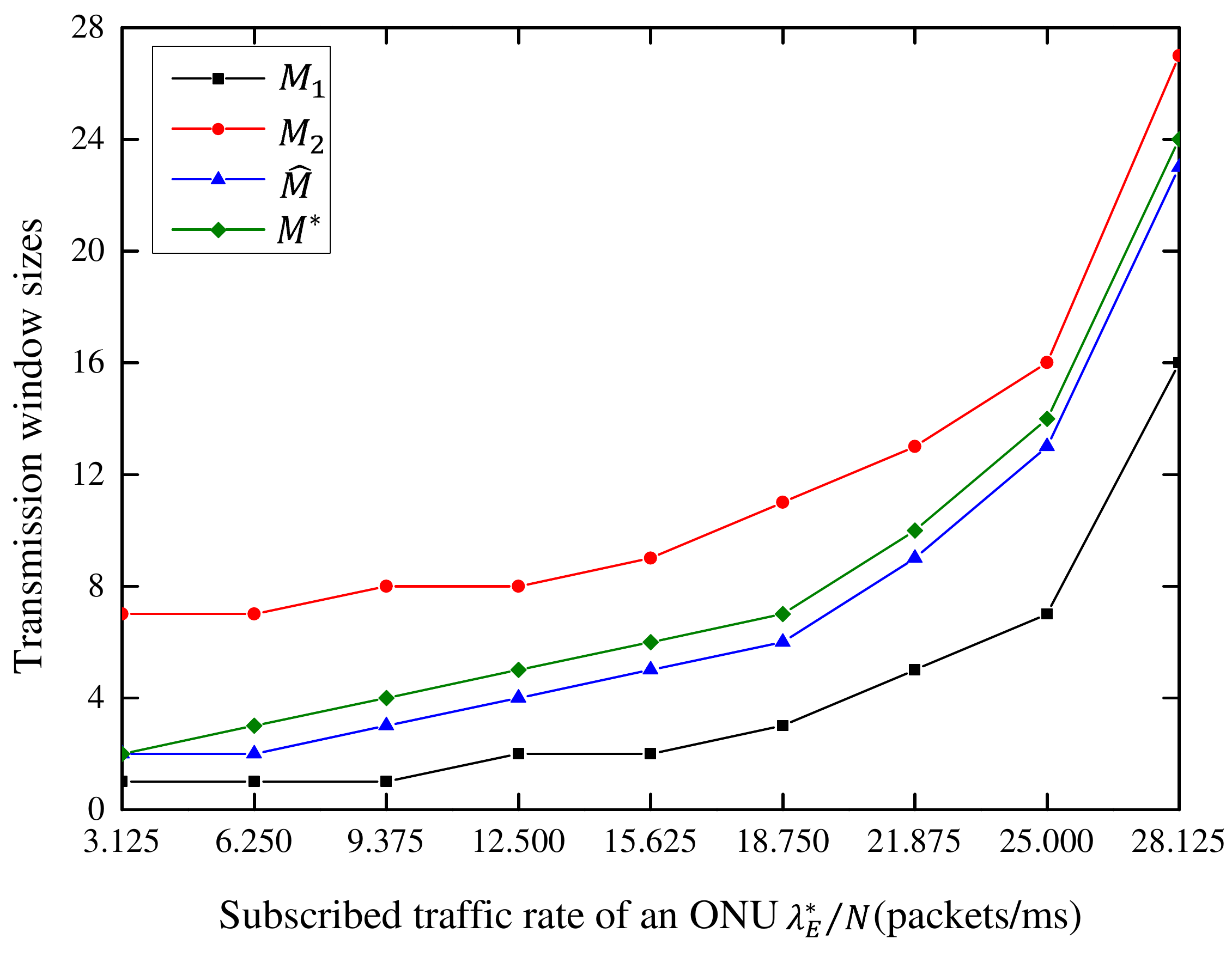}}
\caption{TW sizes vary with tail bound $\varepsilon$ and subscribed traffic rate $\lambda_E^*/N$ respectively: (a) TW sizes vs.$~\varepsilon$ where $\lambda_E^*/N=21.875$ packets/ms and (b) TW sizes vs. $\lambda_E^*/N$ where $\varepsilon=0.05$.}
\end{figure}

Fig.~\ref{fig7-a} and \ref{fig7-b} respectively illustrate that the optimum TW size $M^*$, its lower bound $M_1$, upper bound $M_2$, and approximation $\hat{M}$, vary with the tail bound $\varepsilon$ and the subscribed traffic rate $\lambda_E^*/N$ of each ONU. In these figures, we find that both approximate and optimum TW sizes, $\hat{M}$ and $M^*$, are always bounded between $M_1$ and $M_2$. Moreover, the approximation $\hat{M}$ is uniformly smaller than $M^*$, which can be readily seen from the distributions illustrated in Fig.~\ref{fig6}. The convergence rate of normal distribution $q_n^{\prime}$ is faster than that of $q_n$, thus a smaller TW size is needed to achieve the same probability of tail distribution. In spite of that, as Fig.~\ref{fig7-a} shows, the difference between $\hat{M}$ and $M^*$ is very small in the region $\varepsilon\in[0.001,0.1]$ of our interest in practice. Besides, the selection of TW size is very sensitive to the traffic rate subscribed by a user, as illustrated in Fig.~\ref{fig7-b}, with the growth of $\lambda_E^*/N$, the TW size also increases greatly.

\begin{figure}[!t]
\centering
\includegraphics[width=0.45\textwidth]{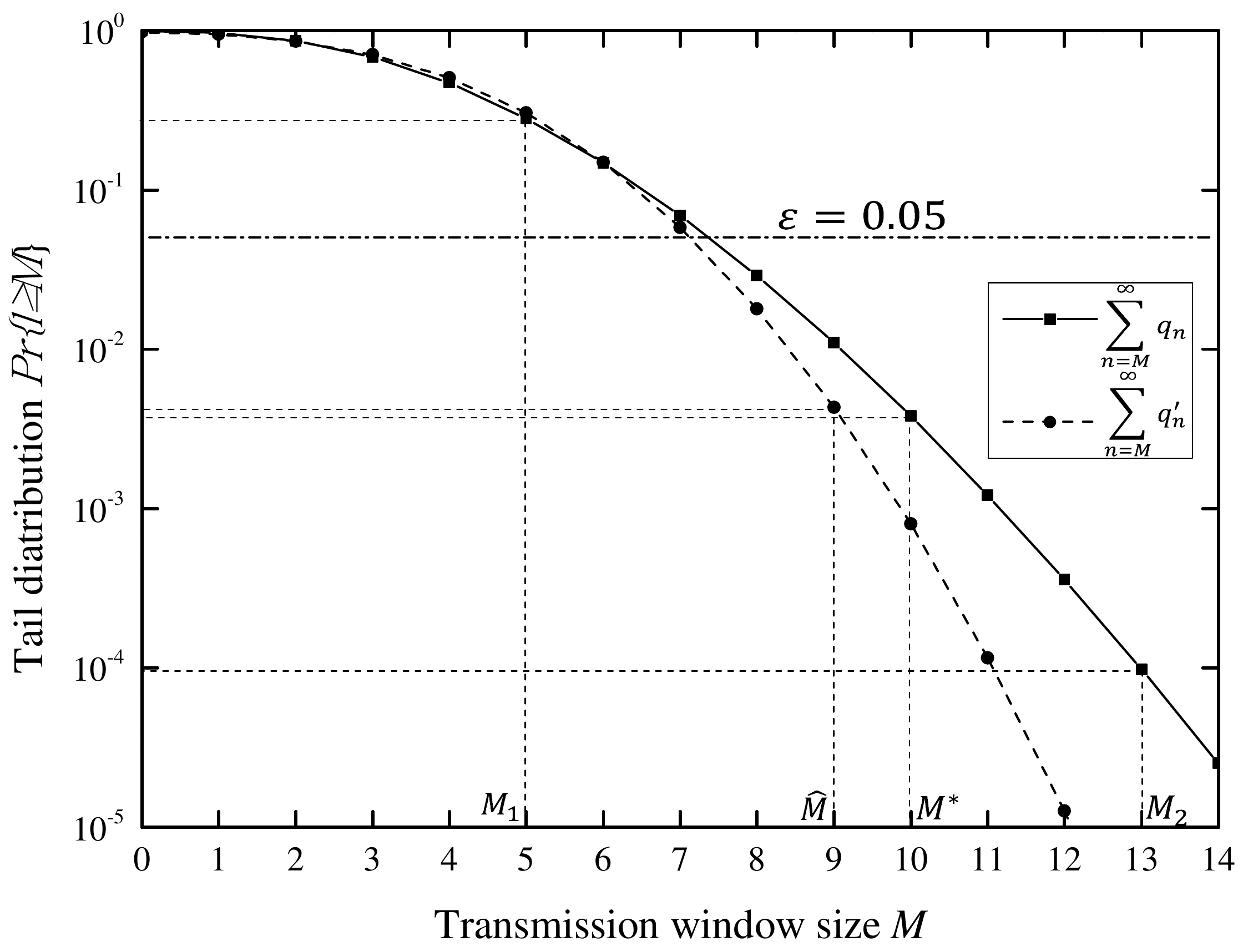}
\caption{Tail distributions of $q_n$ and $q_n^{\prime}$ with $\lambda=21.875$ packets/ms.}
\label{fig8}
\end{figure}

The tail distribution of queue length, $\sum_{n=M}^{\infty}q_n$, and that of approximation, $\sum_{n=M}^{\infty}q_n^{\prime}$, are plotted in Fig.~\ref{fig8} where each ONU inputs traffic with the rate of 21.875 packets/ms. In our interested region $\varepsilon\in[0.001,0.1]$, the difference between the TW sizes selected by tail distributions of $q_n$ and $q_n^{\prime}$ is quite small. When their gap is large, such as in the area $\varepsilon\in[10^{-5},10^{-4}]$, $\varepsilon$ would be extremely small and far below the region of our interest in the practical operation of EPON.

For a fixed $\varepsilon=0.05$, as Fig.~\ref{fig8} shows, despite that $\hat{M}<M^*$, the probability $Pr\{l{\geq}\hat{M}\}$ is still below $\varepsilon$, which means the approximation $\hat{M}$ also satisfies the criterion (\ref{IV-cretirion-org}). If the TW size is set equal to the lower bound $M_1$, the criterion $Pr\{l{\geq}M\}{\leq}\varepsilon$ could be violated and packets may experience longer delay than expected. On the other hand, if the TW size is set equal to the upper bound $M_2$, the criterion can be easily satisfied because $Pr\{l{\geq}M_2\}$ is negligible in comparison with $\varepsilon$. However, the upper bound $M_2$ would be too large to be an effective constraint on malicious users. As a compromise, the approximation $\hat{M}$ can serve as a practical TW size for EPONs.

\subsection{Stability and Delay Performance of EPON}
In this subsection, we study the delay performance of disciplined ONUs in a regular gated-limited service EPON with the TW size limit $M$ given by (\ref{appro-TW}). The gated service discipline is a special case of the gated-limited service discipline with infinite TW size, thus the mean waiting time in gated service is the lower bound of that in gated-limited service.

The EPON system with gated service is stable if the offered load $\rho$ of each ONU is less than $1/N$, i.e. $\lambda<1/\left(N\overline{X}\right)$, which guarantees that input packets will be transmitted steadily and their mean waiting time, or mean queue length, is bounded. However, a bounded mean queue length is not sufficient to guarantee that a regular EPON with gated-limited service is stable due to the limitation of TW size $M$. From the mean queue length formula (\ref{mean-ql}) of an ONU, the stable condition of the gated-limited service EPON is given by
\begin{equation}
\mu_l=\lambda\mu_C=\lambda\cdot\frac{NG}{1-\rho_E}<M,
\label{stable-condition-1}
\end{equation}
where $\rho_E=N\rho=N\lambda\overline{X}$. After some algebraic manipulation, a stable traffic rate $\lambda$ should be bounded by $\hat{\lambda}$ that is defined as follows:
\begin{equation}
\lambda<\hat{\lambda}=\frac{M}{N\left(M\overline{X}+G\right)}
\label{stable-condition-2}
\end{equation}
It¡¯s evident that when $M\to\infty$, $\hat{\lambda}\to1/\left(N\overline{X}\right)$. Furthermore, the TW size limit $M$ is selected based on criterion (\ref{IV-cretirion-org}), which guarantees a very small probability $\varepsilon$ that the queue length will exceed the limit $M$, this is a much more stringent condition than the stable condition (\ref{stable-condition-1}). Therefore, in a regular EPON with gated-limited service, disciplined ONUs with input traffic rate in the region $\lambda\in\left[0,\lambda_E^*/N\right]$ must be all stable, which implies $\lambda_E^*/N{\leq}\hat{\lambda}$.

\begin{figure}[!t]
\centering
\subfigure[]{
\label{fig9-a}
\includegraphics[width=0.45\textwidth]{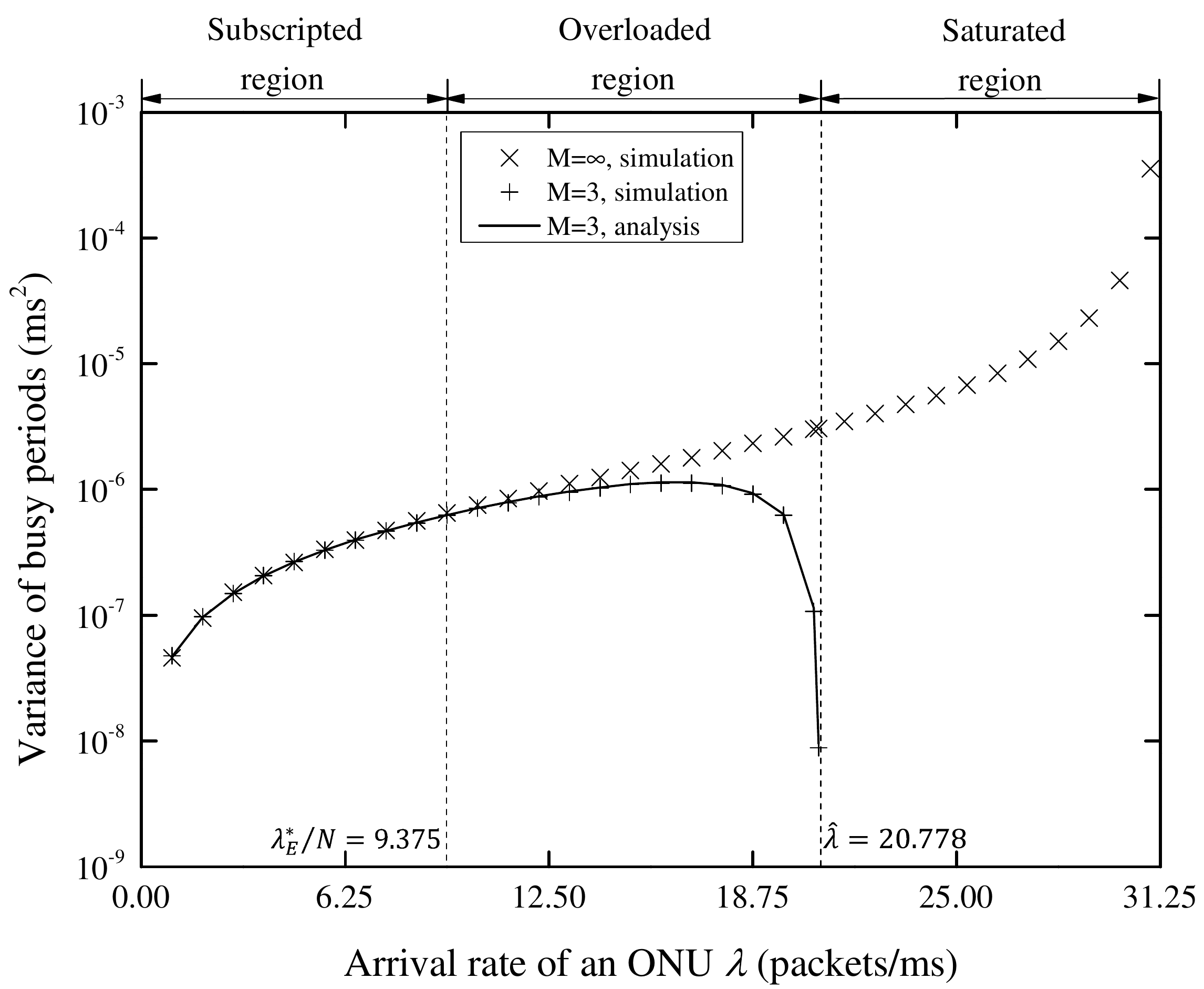}}
\subfigure[]{
\label{fig9-b}
\includegraphics[width=0.45\textwidth]{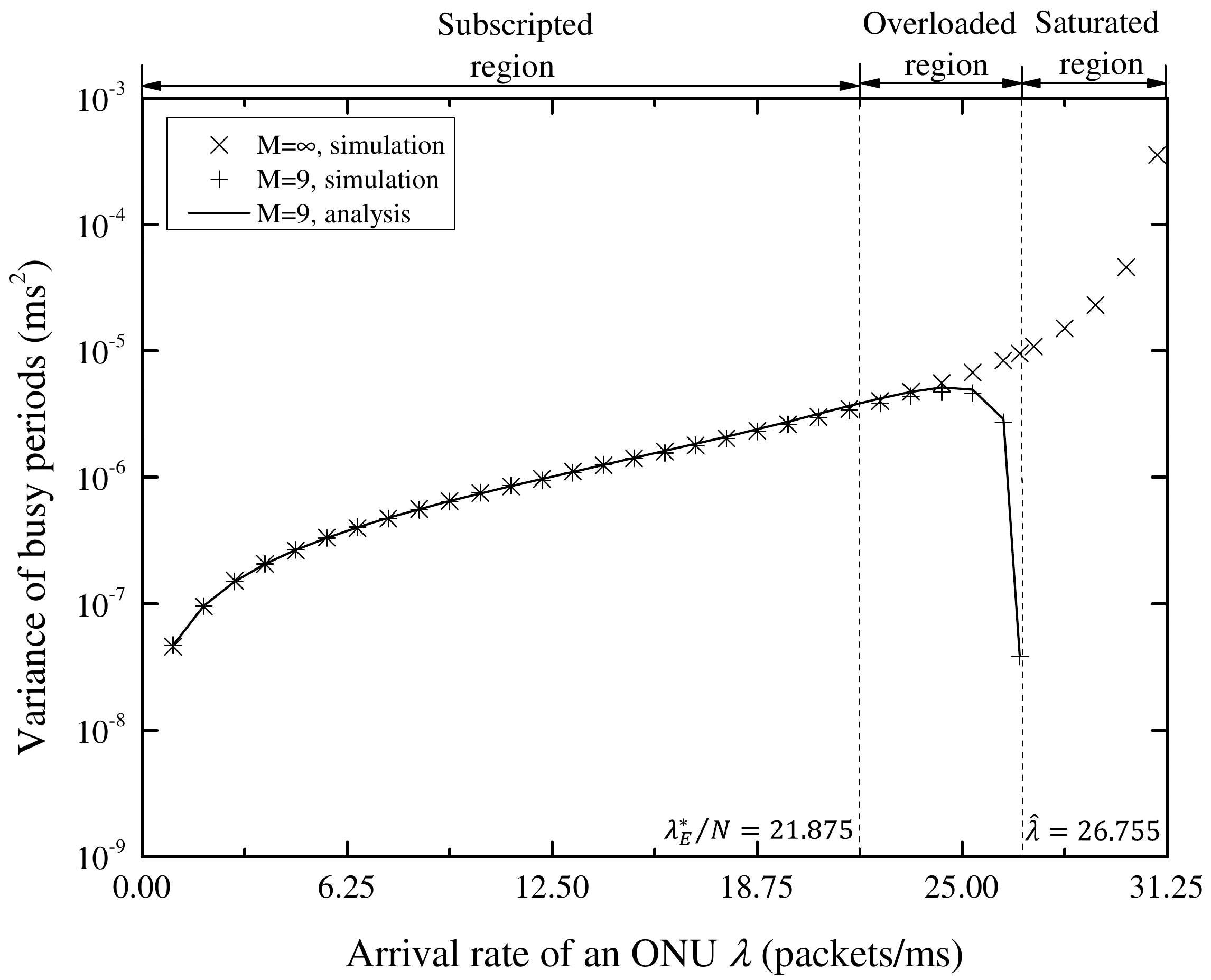}}
\caption{Variance of busy periods under different subscribed traffic rates with $\varepsilon=0.05$: (a) $\lambda_E^*/N=9.375$ packets/ms and (b) $\lambda_E^*/N=21.875$ packets/ms.}
\end{figure}

\begin{figure*}[!t]
\centering
\subfigure[]{
\label{fig10-a}
\includegraphics[width=0.45\textwidth]{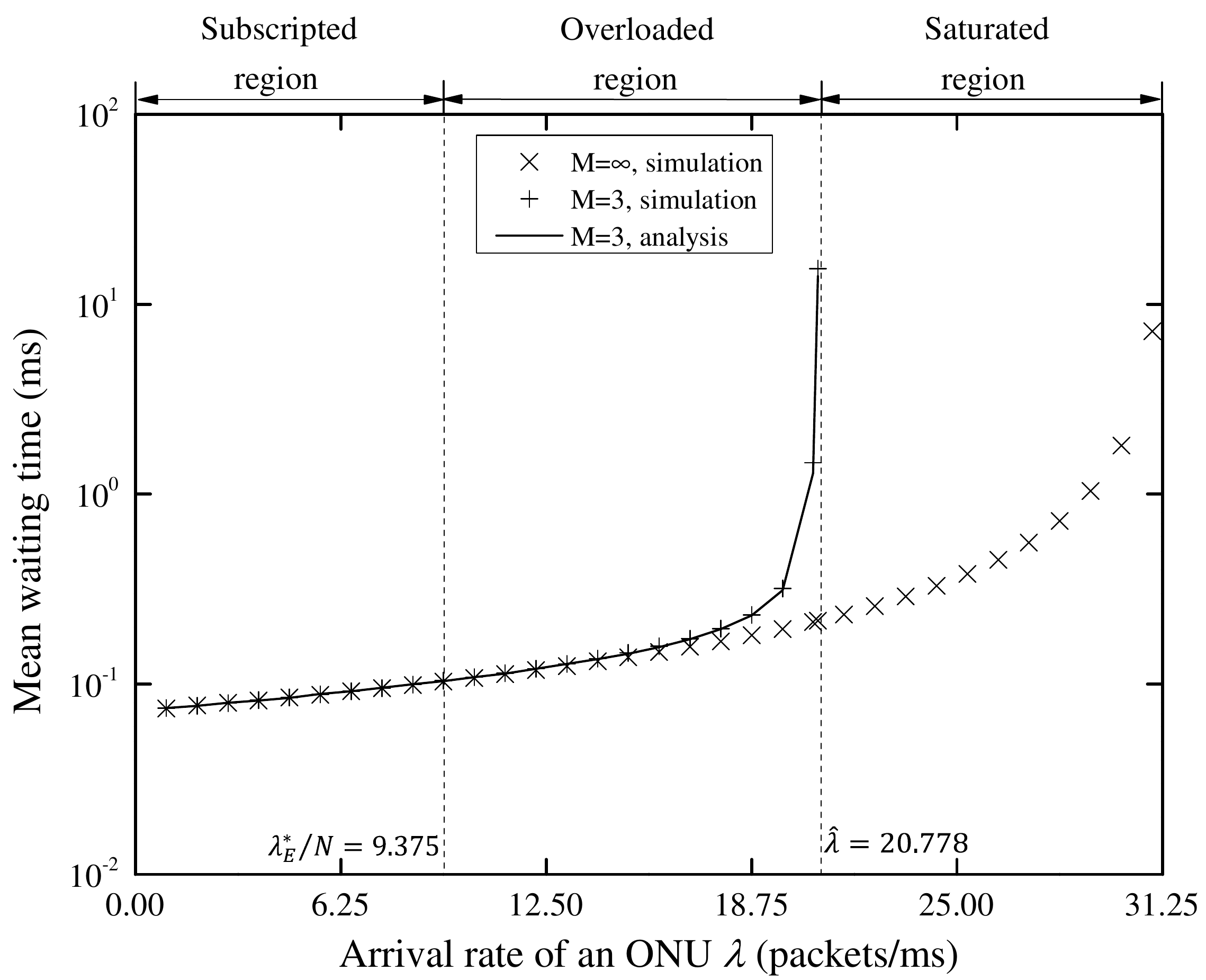}}
\subfigure[]{
\label{fig10-b}
\includegraphics[width=0.45\textwidth]{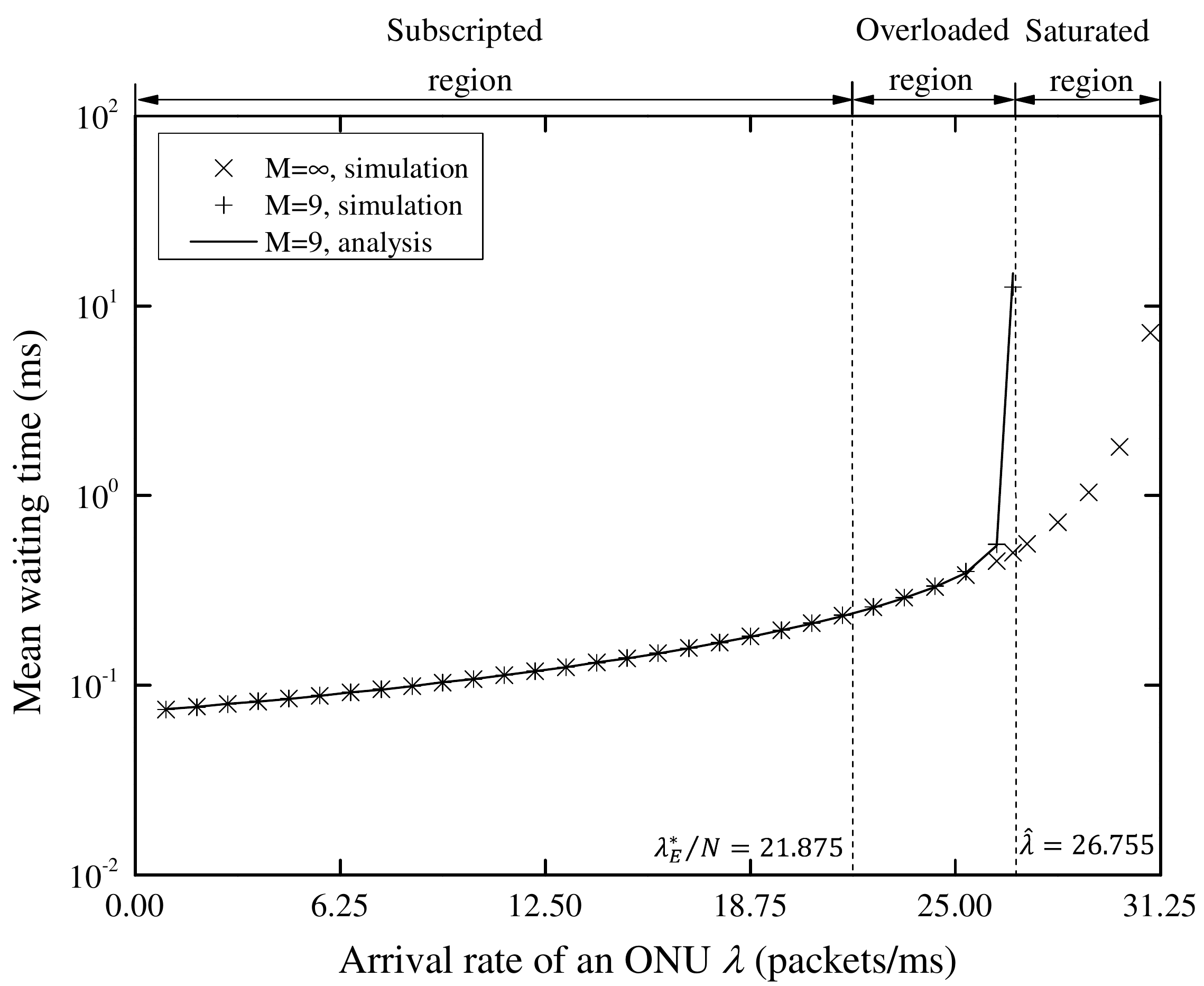}}
\caption{Mean waiting time of an ONU under different subscribed traffic rates with $\varepsilon=0.05$: (a) $\lambda_E^*/N=9.375$ packets/ms and (b) $\lambda_E^*/N=21.875$ packets/ms.}
\end{figure*}

According to the conditions described above, the performance of an ONU in a gated-limited service EPON can be characterized in the following three traffic regions:
\begin{enumerate}[{1)}]
\item
Subscripted region $\lambda\in\left[0,\lambda_E^*/N\right]$. In the subscripted region, the QoS of each ONU in terms of mean delay is guaranteed by the SLA signed with the network operator.
\item
Overloaded region $\lambda\in\left(\lambda_E^*/N,\hat{\lambda}\right)$. If an ONU inputs the packets with the rate higher than its subscripted rate and in the overloaded region, the mean delay is impaired by the limit of the maximum TW size $M$, but it is still bounded. This region provides an adjustment period for the ONU to decrease its input traffic rate when the user experiences a larger than expected delay.
\item
Saturated region $\lambda\in\left[\hat{\lambda},\frac{1}{N\overline{X}}\right)$. In the saturated region, the arrival rate is too high for the OLT to handle. The ONU is unstable when the queue length outside the gate of the buffer becomes unbounded.
\end{enumerate}

In the subscripted region $\lambda\in\left[0,\lambda_E^*/N\right]$ or in the overloaded region $\lambda\in\left(\lambda_E^*/N,\hat{\lambda}\right)$, the mean waiting time of an ONU can be calculated by the procedure described in APPENDIX A. In the saturated region $\lambda\in\left[\hat{\lambda},\frac{1}{N\overline{X}}\right)$, the mean waiting time is unbounded.

The analytical results of mean waiting time in these traffic regions are verified by simulations. We consider a 1G EPON with $N=32$ statistically identical ONUs, and assume that they have signed the same SLAs. The parameters $G=1.512{\mu}s$, $\overline{X}=1{\mu}s$ and $\overline{X^2}=1{{\mu}s}^2$ are the same as those used in Section \uppercase\expandafter{\romannumeral4}-B. Thus the service capacity of the EPON is 1000 packets/ms, evenly divided into 31.25 packets/ms for each ONU. Two scenarios are considered in our study, where each user subscribes to a low traffic rate of 9.375 packets/ms and a high traffic rate of 21.875 packets/ms respectively. According to formula (\ref{appro-TW}), for a fixed $\varepsilon=0.05$, we should set the maximum TW size $M$ equal to 3 and 9 in respect to the above two scenarios.

Fig.~\ref{fig9-a} and~\ref{fig9-b} illustrate the variance of busy periods for each ONU. It is evident that the analytical result given in APPENDIX A is consistent with the simulation result, which validates the accuracy of our analysis in Section \uppercase\expandafter{\romannumeral3}. If we adopt the gated service discipline (i.e., $M$ is infinite), Fig.~9 shows that the variance of busy periods monotonically increases with arrival rate up to infinity. However, with the gated-limited service discipline (i.e., $M$ is finite), the variance of busy periods approaches zero, because each ONU transmits a constant number of $M$ packets in each busy period when the arrival rate is high.

In the subscripted region, as shown in Fig.~\ref{fig10-a} and~\ref{fig10-b}, the disciplined users in the gated-limited service EPON experience the same mean waiting time as that in gated service EPON. This desirable property is due to the criterion $Pr\{l{\geq}M\}{\leq}\varepsilon$ of selecting the TW size $¦¬$, which is sufficiently large to empty the buffered packets almost in every busy period.

In the overloaded region, the ONU will suffer a larger mean delay than expected, which serves as a precaution measure for the ONU to reduce the loading back to the subscripted region. If the ONU continues increasing the input traffic rate to the saturated region, its mean delay tends to infinity, and the service is collapsed to prevent its malicious behavior from impacting the QoS of other disciplined users.

\section{Conclusion}
In this paper, we consider an EPON with gated-limited service discipline as a polling system. Each ONU of an EPON is modeled as an M/G/1 queue with vacations and gated-limited service. A distinguished feature of this model is that there are two queues in the buffer of each ONU: one queue is inside the gate and the other one outside the gate. We extend the traditional geometric approach to derive the Pollaczek-Khinchine formula of mean waiting time. Moreover, the Chernoff bound method is applied to the selection of the optimum TW size. The criterion of selecting the TW size $M$ is to guarantee that the delay performances experienced by disciplined users are bounded, and to constrain malicious users from monopolizing the transmission channel. For this purpose, we devise a simple rule to determine a proper optimum TW size for each ONU of the gated-limited service EPON based on their SLAs.

\begin{appendices}
\section{iterative procedure of calculating mean waiting time $\overline{W}$ and variance of busy periods $\sigma_B^2$}
As analyzed in Section \uppercase\expandafter{\romannumeral3}, it is critical to obtain the second moment of the number of packets served in a busy period $\overline{K^2}$ when calculating the mean waiting time and variance of busy periods. However, the value of $\overline{K^2}$ and that of distribution $q_n~(n=0,1,\cdots,M-1)$ depend on each other, and we can only solve them numerically.

According to Rouche's theorem, the denominator of (\ref{genfun-qn}) has $M$ zeros inside and on $|z|=1$, one of them is $z=1$. Then by Lagrange's theorem \cite{TakagiQueueingBook}, the other $(M-1)$ zeros inside $|z|=1$ are given by
\begin{equation}
z_m=\sum_{n=1}^{\infty}\left.\frac{e^{2{\pi}mni/M}}{n!}\frac{d^{n-1}}{dz^{n-1}}{\Big[H\big(z\big)\Big]}^{n/M}\right|_{z=0},
\label{appendixA-roots}
\end{equation}
for $m=1,2,\cdots,M-1$. Since $Q(z)$ is analytic in $|z|{\leq}1$, the numerator of (\ref{genfun-qn}) must also be zero at $z=z_m$. Therefore, $q_n~(n=0,1,\cdots,M-1)$ satisfy the following $(M-1)$ linear equations:
\begin{equation}
\sum_{n=0}^{M-1}q_n\left(z_m^M-z_m^n\right)=0,~~m=1,2,\cdots,M-1.
\label{appendixA-lineareqs}
\end{equation}
Another equation is given as follows by the condition $Q(1)=1$:
\begin{equation}
\sum_{n=0}^{M-1}q_n\left(M-n\right)=M-\lambda\mu_C=M-\frac{\lambda_EG}{1-\rho_E}.
\label{appendixA-normalized-condition}
\end{equation}
Thus, if we know the expression of $H(z)$, we can solve $q_n~(n=0,1,\cdots,M-1)$ by combining (\ref{appendixA-roots})-(\ref{appendixA-normalized-condition}), then obtain $\overline{K^2}$ based on (\ref{K2-qk}), which is
\begin{equation*}
\overline{K^2}=\sum_{n=0}^{M-1}n^2q_n+M^2\left({1-\sum_{n=0}^{M-1}q_n}\right).
\end{equation*}

However, the expression of $H(z)$ is instead dependent on $\overline{K^2}$. Therefore, given a calculation accuracy $\delta$, we can numerically solve $\overline{K^2}$ through the following iteration procedure:
\begin{enumerate}[{Step 1:}]
\item
 $\overline{K^2}=0$;
\item
Calculate $H(z)$ by combining (\ref{mean-K}), (\ref{genfun-hn})-(\ref{var-cycle});
\item
Solve $z_m,m=1,2,\cdots,M-1$ by (\ref{appendixA-roots});
\item
Solve $q_n,n=0,1,\cdots,M-1$ by combining (\ref{appendixA-lineareqs}) and (\ref{appendixA-normalized-condition});
\item
If $\left|\sum_{n=0}^{M-1}n^2 q_n+M^2\left(1-\sum_{n=0}^{M-1}q_n\right)-\overline{K^2}\right|>\delta$, $\overline{K^2}=\sum_{n=0}^{M-1}n^2q_n+M^2\left(1-\sum_{n=0}^{M-1}q_n\right)$, go to Step 2;
\item
Output $\overline{K^2}$.
\end{enumerate}
Then, we can easily obtain the variance of busy periods for an ONU by substituting $\overline{K^2}$ and (\ref{mean-K}) into (\ref{busyperiod-var}), and the mean waiting time by combining (\ref{III-A-Theorem1}), (\ref{mean-ONU-vacation}), (\ref{secmoment-ONUvacation-2}), (\ref{mean-K}) and $\overline{K^2}$.

\section{proof of theorem 2}
Define the following function:
\begin{equation}
f\left(t,z\right)\!=\!exp\Big[-\big(\mu_l+t\big){\log}z\!+\!\lambda\mu_C\big(z-1\big)\!+\!\frac{1}{2}{\lambda}^2\sigma_C^2{\big(z-1\big)}^2\Big],
\label{appendixB-1}
\end{equation}
where $z>1$ and $t{\geq}0$. We know that the following inequality holds for all $x{\geq}0$:
\begin{equation}
-x{\leq}-{\log}\left(1+x\right){\leq}-\left(x-\frac{1}{2}x^2\right).
\label{appendixB-2}
\end{equation}
Let $x=z-1$, and apply (\ref{appendixB-2}) to (\ref{appendixB-1}), then we have the following inequality:
\begin{equation}
f_1(t,z){\leq}f(t,z){\leq}f_2(t,z),
\label{appendixB-3}
\end{equation}
where the two functions $f_1(t,z)$ and $f_2(t,z)$ are defined as follows:
\begin{equation}
f_1(t,z)=exp\Big[-t\big(z-1\big)+\frac{1}{2}{\lambda}^2\sigma_C^2{\big(z-1\big)}^2\Big],
\label{appendixB-4}
\end{equation}
and
\begin{equation}
f_2(t,z)=exp\Big[-t\big(z-1\big)+\frac{1}{2}\big(\sigma_l^2+t\big){\big(z-1\big)}^2\Big].
\label{appendixB-5}
\end{equation}

Take the derivatives of (\ref{appendixB-1}), (\ref{appendixB-4}) and (\ref{appendixB-5}), we obtain
\begin{align}
-\frac{\mu_l+t}{z^*}+\lambda\mu_C+{\lambda}^2\sigma_C^2\left(z^*-1\right)&=0, \label{appendixB-6}\\
-t+{\lambda}^2\sigma_C^2\left(z_1-1\right)&=0,
\label{appendixB-7}
\end{align}
and
\begin{equation}
-t+\left(\sigma_l^2+t\right)\left(z_2-1\right)=0.
\label{appendixB-8}
\end{equation}
It follows from (\ref{appendixB-3}), the following inequalities should hold:
\begin{align}
\inf_{z>1}f_1(t,z)&=f_1(t,z_1){\leq}f_1(t,z^*){\leq}f(t,z^*), \label{appendixB-9} \\
\inf_{z>1}f(t,z)&=f(t,z^*){\leq}f(t,z_2){\leq}f_2(t,z_2).
\label{appendixB-10}
\end{align}
Combining (\ref{appendixB-9}) and (\ref{appendixB-10}), the following expression can be obtained from $z_1$ and $z_2$ given by (\ref{appendixB-7}) and (\ref{appendixB-8}) respectively,
\begin{align}
f_1(t,z_1)&=exp\left[-\frac{t^2}{2{\lambda}^2\sigma_C^2}\right]{\leq}f(t,z^*)  \nonumber\\
&{\leq}exp\left[-\frac{t^2}{2\left(\sigma_l^2+t\right)}\right]=f_2(t,z_2).
\label{appendixB-11}
\end{align}

Let $t^*$, $t_1$ and $t_2$ be the solutions that respectively satisfy the following three equations:
\begin{equation*}
exp\left[-\frac{t_1^2}{2{\lambda}^2\sigma_C^2}\right]=f(t^*,z^*)=exp\left[-\frac{t_2^2}{2\left(\sigma_l^2+t_2\right)}\right]=\varepsilon.
\end{equation*}
Then, according to (\ref{appendixB-11}), we have
\begin{align}
&exp\left[-\frac{{t^*}^2}{2{\lambda}^2\sigma_C^2}\right]{\leq}f(t^*,z^*)=exp\left[-\frac{t_1^2}{2{\lambda}^2\sigma_C^2}\right]  \nonumber\\
&=exp\left[-\frac{t_2^2}{2\left(\sigma_l^2+t_2\right)}\right]{\leq}exp\left[-\frac{{t^*}^2}{2\left(\sigma_l^2+t^*\right)}\right].
\label{appendixB-12}
\end{align}
Since those exponential functions in (\ref{appendixB-12}) are monotonically decreasing with $t$, we have
\begin{equation}
t_1{\leq}t^*{\leq}t_2.
\label{appendixB-13}
\end{equation}
Substituting $t=M-\mu_l$ into (\ref{appendixB-13}), we obtain
\begin{equation}
M_1{\leq}M^*{\leq}M_2.
\label{appendixB-Mrelation}
\end{equation}
Hence, the smallest integer $M_1$ that satisfies the following inequality is a lower bound of $M^*$:
\begin{equation*}
exp\left[-\frac{{\left(M_1-\mu_l\right)}^2}{2{\lambda}^2\sigma_C^2}\right]{\leq}\varepsilon,
\end{equation*}
and it can be explicitly expressed as follows:
\begin{equation}
M_1=\left\lceil\mu_l+\lambda\sigma_C\sqrt{2\log\varepsilon^{-1}}\right\rceil=\left\lceil\mu_l+\lambda\sigma_C\sqrt{2\alpha}\right\rceil.
\label{appendixB-14}
\end{equation}
Similarly, the smallest integer $M_2$ that satisfies the following inequality is an upper bound of $M^*$:
\begin{equation*}
exp\left[-\frac{{\left(M_2-\mu_l\right)}^2}{2\left({\lambda}^2\sigma_C^2+M_2\right)}\right]{\leq}\varepsilon,
\end{equation*}
and it can be given as follows:
\begin{align}
M_2&=\left\lceil\mu_l+\log\varepsilon^{-1}\!+\!\sqrt{{\left(\log\varepsilon^{-1}\right)}^2\!+\!2\log\varepsilon^{-1}\left(\mu_l+\lambda^2\sigma_C^2\right)}\right\rceil  \nonumber \\
   &=\left\lceil\mu_l+\alpha+\sqrt{{\alpha}^2+2\alpha\sigma_l^2}\right\rceil.
\label{appendixB-15}
\end{align}
We obtain (\ref{IV-B-theorem2}) by combining (\ref{appendixB-Mrelation})-(\ref{appendixB-15}).
\end{appendices}

\bibliographystyle{IEEEtran}
\bibliography{IEEEabrv,IEEERef}
\end{document}